\documentclass[aps,prd,reprint,superscriptaddress,amsmath,amssymb,nofootinbib,floatfix]{revtex4-2}
\pdfoutput=1

\AtBeginDocument{%
  \setlength{\abovedisplayskip}{7pt plus 2pt minus 2pt}%
  \setlength{\belowdisplayskip}{7pt plus 2pt minus 2pt}%
  \setlength{\abovedisplayshortskip}{4pt plus 1pt minus 1pt}%
  \setlength{\belowdisplayshortskip}{5pt plus 1pt minus 1pt}%
  \setcounter{topnumber}{3}%
  \setcounter{bottomnumber}{3}%
  \setcounter{totalnumber}{5}}
\usepackage[T1]{fontenc}
\usepackage[utf8]{inputenc}
\usepackage{booktabs}
\usepackage{multirow}
\usepackage{graphicx}
\usepackage{xcolor}
\usepackage{hyperref}

\begin{document}

\title{The efficiency gap of two-step forecasts for stochastic gravitational wave background searches}

\author{Bo-Qiang~Lu}
\email{bqlu@huznu.edu.cn}
\affiliation{School of Science, Huzhou Normal University, Huzhou, Zhejiang 313000, P. R. China}

\begin{abstract}
The stochastic gravitational-wave background is targeted by the millihertz observatories LISA,
Taiji and TianQin, and its forecast precision rests on a Whittle approximation to the
periodogram likelihood.  The quoted scale is that of a joint fit a two-step forecast need not
reach.  We show that the residual risk of a two-step forecast lies in the estimator route
rather than in the likelihood.  At the exact Whittle form we separate the forecast error into a
deterministic bias $\mathbf{b}$ and an efficiency gap $\mathbf{C}_b$, both in closed form.  
We find that this gap is governed by the span of the first-step nuisance fit. It decreases 
substantially as the span increases and vanishes when the first-step fit spans the entire band 
or when the response exhibits no within-bin structure. The gap does not arise as a cost of the prior.
Across the LISA band the gap spans more than an order of magnitude, the amplitude and tilt inflation
falling from about thirty and twenty to unity at a band-wide first step.  A foreground from the
released LDC2a population moves the mid-band gap by an order of magnitude.  A two-step forecast
should therefore report the span of its first-step fit with its intervals, since that span
decides whether they sit at the Fisher scale or an order of magnitude above it.
\end{abstract}


\maketitle
\section{Introduction}

A stochastic gravitational-wave background (SGWB) is anticipated from the superposition of
numerous unresolved sources spanning cosmological and astrophysical epochs.  The cosmological
component includes the tensor perturbations seeded by inflation, the radiation from first-order
phase transitions and the emission of topological defects such as cosmic strings and domain
walls \cite{Caprini:2018,Caprini:2019,Hindmarsh:2014}, while unresolved compact-binary
coalescences dominate the astrophysical one \cite{Romano:2017,Karnesis:2021}.  Ground-based
interferometers have opened the gravitational-wave window \cite{Abbott:2016}, and the
forthcoming space-based observatories, the Laser Interferometer Space Antenna (LISA)
\cite{LISA:2017}, Taiji \cite{Taiji:2017} and TianQin \cite{TianQin:2016}, will extend
stochastic searches across several decades in the millihertz band \cite{Berti:2022}.  Because
the SGWB encodes energy scales and eras beyond the reach of electromagnetic and collider
probes, it constrains early-Universe cosmology and the evolution of compact sources
\cite{Caprini:2018,Romano:2017}.

Forecasting the achievable parameter precision reduces in practice to the Fisher information
matrix of a Whittle (Gaussian) approximation to the periodogram
likelihood~\cite{Whittle:1952,Whittle:1953,Romano:2017}.  The resulting error bars depend both
on that approximation and on how the likelihood is realised as an estimator of the band-shared
nuisance.  The Whittle likelihood treats each periodogram
$\hat C_k(f) = |\tilde d_k(f)|^2/T_{\rm seg}$ as an exponential draw
$\sim P(f)\,\chi^2_{2N_s}/(2N_s)$, exact only as $N_s \to \infty$.  At finite segmentation the
maximum-likelihood estimator $\hat\theta$ is biased at order $1/N_s$.
Ref.~\cite{Franciolini:2025PRD} have performed the first systematic study of these biases.  For
LISA and Taiji with the standard $11.5\,$d segmentation they document a bias-dominated
single-parameter Gaussian diagnostic whose bias-to-statistics ratio reaches
$\mathcal{B}/\mathcal{S}=62.5$ (Sec.~\ref{sec:biasdiag}).  That ratio pertains to the per-bin
level estimator, and whether it reaches a \emph{fitted} template parameter depends on the
estimator route.  The per-bin biased object enters the fit as data in a two-step implementation,
whereas a single joint maximisation of the same objective leaves the fitted parameter set
unbiased.

What a forecast therefore has to settle is whether the likelihood matters and whether the route
does.  The likelihood question has a sharp answer.  The periodogram likelihoods that treat the
periodogram as Gaussian fall into a one-parameter family whose per-bin objectives differ by a term
proportional to $\ln P$.  The coefficient of that term moves the per-bin maximum away from the
sample mean, but only along a deterministic multiple of it, $\hat P_c=f_c(N_s)D$, so the
logarithmic variance $\mathrm{Var}[\ln\hat P_c]=\psi'(N_s)$ is the same for every member and the
estimator efficiency is invariant across the family.  The deterministic bias behaves differently,
moving with the position of the per-bin objective relative to the sample mean.  The log-normal and
Hamimeche--Lewis forms~\cite{Hamimeche:2008} lie outside the family, so the invariance is a
statement about the Gaussian likelihoods in use.  Section~\ref{sec:gapblind} establishes it, which
lets the present analysis fix the likelihood at one member without loss of generality for the
efficiency.

The second question is the subject of this work.  A forecast realised in two steps is
inefficient relative to a single joint fit.  The first step fits the band-shared nuisance
parameters, and the second fits the signal template with the nuisance held at its first-step
value.  The inefficiency is a property of the estimator route alone.  Such a two-step
realisation is the plug-in construction of the estimation literature, where a sequential
estimator is analysed as a member of a moment-based class with its asymptotic covariance
derived accordingly \cite{Newey:1984}, under general conditions for consistency, efficiency
and correct inference \cite{Pagan:1986}.  The conditions for a two-step estimator to attain
the joint efficiency bound are also known \cite{Ackerberg:2014}, the reference being the
Fisher matrix marginalised over a Gaussian nuisance through its Schur complement
\cite{Taylor:2010}.  The same partition into a nuisance and a signal component, with the second
returning nothing to the first, is the cutting of feedback that modular Bayesian inference uses
to keep one component of a model from influencing another~\cite{Frazier:2025}.

We separate the error of a two-step forecast into a bias vector $\mathbf{b}$ and an efficiency gap
$\mathbf{C}_b \equiv \mathbf{C}_{\hat\theta}-\mathbf{F}^{-1}$, the two quantities it reports beyond
its Fisher scale, through the mean-squared-error decomposition
$\mathbf{\Sigma}_{\rm MSE}=\mathbf{F}^{-1}+\mathbf{C}_b+\mathbf{b}\mathbf{b}^{\rm T}$ with
$\mathbf{b}$ the deterministic point bias.  Both are calibrated here in closed form from the block
information matrix of the levels and the shared nuisance rather than from a Monte-Carlo, and
checked against an explicit posterior sample of the same model.

The closed form identifies the quantity that controls the gap.  A single bin constrains the
shared nuisance only along the projection of its own nuisance response, the remaining directions
falling to the prior, whereas the levels respond to every direction.  The information that the
joint route borrows from the other bins is therefore carried by the \emph{within-bin shape} of
the nuisance response, and the gap vanishes identically when that shape is absent.  On the LISA
band, with $N_k=20$ bins and a twenty-per-cent Gaussian prior on the noise amplitudes, the gap
spans more than an order of magnitude across the band, from single figures to above a hundred
(Table~\ref{tab:cbprofile}).  The fitted amplitude and tilt carry a $1\sigma$ inflation of
order thirty and twenty, respectively, when the first step runs in every bin separately
(Table~\ref{tab:ngswitch}).  Widening the first-step span lowers the inflation monotonically to
exactly unity, the internal null of this analysis.

The same span controls what a reconstruction of $\Omega_{\rm GW}(f)$ inherits.  A binned
reconstruction fits the levels themselves rather than a single template across the band, so the
gap enters the reconstructed spectrum directly.  The per-bin level error is inflated by a factor
of a few to above ten, a shape-isolating reconstruction by an order of magnitude more, and
band-averaging dilutes only the statistical term.  A Gaussian prior misplaced by ten percent of
its width, as a mis-specified instrument model would leave it, shifts the two-step levels by up
to $10^{-1}$ in fractional terms against $10^{-4}$ for the joint levels.  The bias and efficiency
channels are therefore controlled by the same route.

The operational forecasts that motivate this analysis live on the likelihood axis.  The binned
millihertz reconstruction of Refs.~\cite{Flauger:2021,Kume:2025} uses a Gaussian likelihood of
the same quadratic form in the residual $(D-P)/P$ as the primitive of
Appendix~\ref{app:likes}, without a log-determinant term.  Alternative objectives appear in the
same class, among them the log-normal form whose skewness correction removes the bias from the
non-Gaussianity of the unaveraged periodogram~\cite{Hamimeche:2008}.  The ground-based
cross-correlation search lies outside the class, its binned statistic approximately sufficient
and its variance taken from the measured autopower, so no model-dependent determinant enters the
first stage~\cite{Matas:2021}.  These are choices of objective.  The inflation calibrated here
belongs to the realisation, and Section~\ref{sec:gapblind} separates the two.  Within this
family, changing the working likelihood leaves the estimator efficiency untouched, whereas
changing the span of the first step does not.

Current SGWB forecasts~\cite{Flauger:2021,Martinovic:2021,Gowling:2021,Smith:2019} and
cross-correlation methodology~\cite{Allen:1999,Matas:2021} report Whittle Fisher
uncertainties, and instrument-calibration studies~\cite{Capote:2024} fix the noise-model inputs
those forecasts assume.  A space-based interferometer observes continuously with overlapping
sources, so it offers no quiet stretch from which to estimate the spectrum, and the spectrum
entering a Whittle likelihood is obtained on source, its accuracy governing both the bias of the
recovered parameters and the width of their intervals~\cite{Aimen:2025}.  The closed form is
written for a general template and calibrated on the power law, the family in which the
millihertz forecasts are conventionally quoted.  At the parameter values adopted here, the broken
and smooth broken power laws move the per-bin factor by up to a factor of two while leaving the
projected parameter inflation within about a tenth of its power-law value
(Sec.~\ref{sec:calibration}).

\begin{figure*}[tp]
  \centering
  \includegraphics[width=0.90\linewidth]{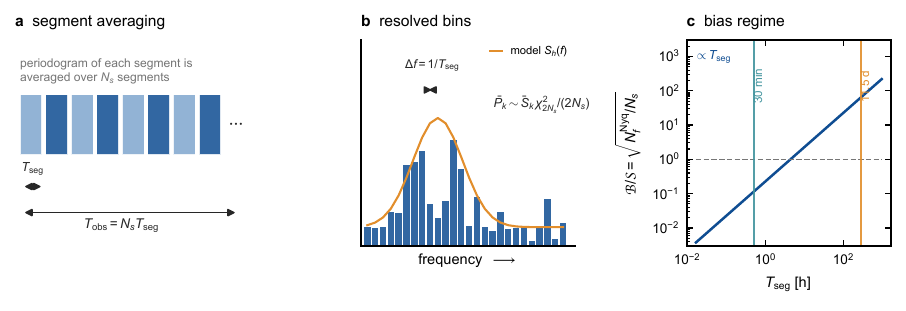}
  \caption{Schematic of the segmentation and of the two channels of estimator inefficiency.
  \textbf{a}, $T_{\rm obs}$ is cut into $N_s$ segments of length $T_{\rm seg}$ whose
  periodograms are averaged.  \textbf{b}, each segment resolves $N_f$ bins of width
  $\Delta f=1/T_{\rm seg}$ [Eq.~\eqref{eq:Nf}], so the averaged periodogram is a $\chi^2_{2N_s}$
  variable rather than Gaussian, and each coarse bin of the reconstruction carries $m$ of those
  fine bins.  \textbf{c}, the bias-to-statistics ratio $\mathcal{B}/\mathcal{S}\propto T_{\rm seg}$
  [Eq.~\eqref{eq:BSscaling}] grows linearly with the segment length, and the point-bias
  inflation $R^2-1\propto T_{\rm seg}^2$ [Eq.~\eqref{eq:Rscaling}] grows as its square;
  vertical lines mark the $30\,$min and $11.5\,$d baselines.}
  \label{fig:schematic}
\end{figure*}

Fig.~\ref{fig:schematic} summarises the segmentation geometry and the two dimensionless
quantities that control the two channels.  Section~\ref{sec:setup} fixes the notation, the
likelihood primitives and the templates, and Section~\ref{sec:biasdiag} reviews the
single-parameter diagnostic and its reporting scales.  Section~\ref{sec:gap} derives the
efficiency gap in closed form for a band-shared nuisance, and states its exact endpoints and
zero criterion.  Section~\ref{sec:calibration} calibrates it across the LISA band,
Section~\ref{sec:params} carries the calibration to the fitted parameters and checks it against
an explicit posterior sample (Sec.~\ref{sec:parammc}), Section~\ref{sec:recon} translates it
into the reconstruction of $\Omega_{\rm GW}(f)$, and Section~\ref{sec:vectorbias} treats the
bias-vector channel.  Section~\ref{sec:network} gives the network Fisher forecast with the
adopted overlap-reduction functions and white-dwarf confusion model, both derived in
Appendix~\ref{app:network}, and Section~\ref{sec:benchmark} validates the analysis on synthetic
LISA data.  Section~\ref{sec:discussion} places the calibration among the operational searches,
Section~\ref{sec:conclusions} concludes, and Appendix~\ref{app:likes} collects the likelihood
formulae used throughout.

\section{Setup: likelihoods, templates, detectors}
\label{sec:setup}
\subsection{Likelihoods}
We work in terms of the dimensionless fractional energy density per logarithmic frequency
interval, $\Omega_{\rm GW}(f) \equiv (1/\rho_{\rm c})\,{\rm d}\rho_{\rm GW}/{\rm d}\ln f$.  A
measurement produces complex Fourier amplitudes $\tilde d_k(f)$ for $k = 1, \ldots, N_s$
segments of length $T_{\rm seg}$, with periodogram
$\hat C_k(f) = |\tilde d_k(f)|^2/T_{\rm seg}$, $N_s = T_{\rm obs}/T_{\rm seg}$ and $N_f$
frequency bins of width $\Delta f = 1/T_{\rm seg}$ per segment.  In the Whittle approximation
each $\hat C_k(f)$ is an independent draw from an exponential distribution of mean
$S_h(\theta; f)$, so that $2N_s\,\bar C(f)/S_h(\theta;f)\sim\chi^2_{2N_s}$ for the segment
average, and the joint likelihood of the periodograms is
\begin{align}
  \ln\mathcal{L}_{\rm W}(\theta|D)
    &= \sum_{k=1}^{N_s} \ln\mathcal{L}_{\rm W}\!\left(\theta^k|D^k\right)
       \notag\\
    &= -\sum_{k=1}^{N_s}\sum_{f}
       \left[ \frac{\hat C_k(f)}{S_h(\theta^k; f)}
              + \ln S_h(\theta^k; f) \right],
  \label{eq:whittle}
\end{align}
with $\theta^k \equiv (\theta_h,\theta_n^k)$ carrying the segment
index~\cite{Franciolini:2025PRD}.  The SGWB parameters $\theta_h$ are common to all segments
while the noise parameters $\theta_n^k$ may vary; here the instrumental noise is known and
stationary, so $S_h(\theta^k;f) = S_h(\theta;f)$ and the index is dropped.  The model power
spectral density (PSD)
$S_h(\theta; f) = S_h^{\rm inst}(f) + S_h^{\rm SGWB}(\theta; f) + S_h^{\rm gal}(f)$ sums the
instrumental noise, the SGWB signal and the Galactic confusion, and the energy-density
representation is converted through
\begin{equation}
  S_h^{\rm SGWB}(\theta;f) = \frac{3 H_0^2}{4\pi^2}\,\frac{\Omega_{\rm GW}(\theta;f)}{f^3}\,,
  \label{eq:omega2sh}
\end{equation}
with $H_0 = 67.4\ {\rm km\,s^{-1}\,Mpc^{-1}}$.  Averaging Eq.~\eqref{eq:whittle} over the
segment ensemble, $\mathbb{E}[\hat C(f)] = S_h(\theta;f)$, gives the Fisher matrix used for all
forecasts,
\begin{equation}
  F_{ab} \;=\; -\,\mathbb{E}\!\left[
     \frac{\partial^2 \ln\mathcal{L}_{\rm W}}{\partial\theta_a\partial\theta_b}
     \right]
  \;=\; N_s \sum_f
     \frac{\partial \ln S_h}{\partial\theta_a}
     \frac{\partial \ln S_h}{\partial\theta_b}\,,
  \label{eq:fisherdef}
\end{equation}
the sum running over positive frequencies only.  Each bin already carries the two degrees of
freedom of a complex Fourier coefficient, so no explicit factor of $1/2$ appears (the $N_s/2$
of the per-bin estimator $\hat P^{\rm D}_k$, Table~\ref{alg:twostep}, belongs to that estimator, not
the Fisher).  On the \emph{physical resolved} grid, $\Delta f_k=1/T_{\rm seg}$ and
$N_f=(f_{\max}-f_{\min})T_{\rm seg}=9.93\times10^{4}$ at the $11.5\,$d baseline
[Eq.~\eqref{eq:Nf}], the discrete sum is the Riemann form of the band integral, and at fixed
$T_{\rm obs}$ the Fisher is re-segmentation invariant because $N_f\propto1/N_s$ with
$N_sN_f=T_{\rm obs}(f_{\max}-f_{\rm min})$ fixed by the observation.  A coarsened grid
under-resolves the template shape and returns a correspondingly larger error.  The naive
(bias-uncorrected) uncertainty on $\theta_a$ is the \emph{marginal} Fisher error
$\sigma_{{\rm naive},a}\equiv\sqrt{(\mathbf{F}^{-1})_{aa}}$ at that physical count, against
which every headline $R$ and $R_i$ is quoted.  The two coarser readings, $\sigma^{\rm cov}_{a}$
for a coarsened analysis grid $N_f^{\rm an}$ exceed it by $\sqrt{N_f/N_f^{\rm an}}$, so the $R$
values quoted on that grid are smaller by the same factor in $b/\sigma$ and the physical-grid
equivalents accompany them.

We implement the likelihood primitives collected in Table~\ref{tab:likes} and
Appendix~\ref{app:likes}, which differ only in how the periodogram is treated, namely the
exact Whittle form of Eq.~\eqref{eq:whittle} and the Gaussian, quadratic and
estimator-subtracted approximations in common pipeline use.  Alternative periodogram likelihoods,
in particular the log-normal and Hamimeche--Lewis forms~\cite{Hamimeche:2008}, lie outside the
Gaussian family of Eq.~\eqref{eq:gapblind} and are not exercised here.  The
working likelihood of this work is the exact Whittle form of Eq.~\eqref{eq:whittle}, whose
per-bin estimator $\hat P^{\rm W}_k=D_k$ is unbiased.  Realised as a single joint fit over the band
it defines the exact-Whittle arm against which the contrast arm below is measured.  The
determinant-approximate \texttt{gauss\_D} likelihood of Eq.~\eqref{eq:lD}, with
$D\equiv\hat C_k(f)$ the periodogram of bin $k$ and $P\equiv S_h(\theta;f)$ the model PSD, is
retained as that contrast arm for its closed-form Fisher.  The arm is realised by the composite
two-step estimator of Table~\ref{alg:twostep}, whose inner step is the per-bin
maximum-likelihood estimator of Eq.~\eqref{eq:pD}.  It approximates the \texttt{gauss\_D}
maximiser rather than attaining it, the discarded log-determinant term being discussed in
App.~\ref{app:likes}.  The difference it makes is neither a property of the likelihood nor of
the estimator efficiency, as Section~\ref{sec:gapblind} shows.


\subsection{Templates}
\label{sec:templates}

The forecasts are carried out on families of power-law (PL), broken power-law (BPL) and
smooth broken power-law (SBPL) templates.
\begin{align}
  \Omega_{\rm PL}(f)        &= A\,(f/f_{\rm ref})^{\alpha}, \\
  \Omega_{\rm BPL}(f)       &= A\,\begin{cases}
        (f/f_p)^{\alpha_{\rm in}},  & f \le f_p,\\
        (f/f_p)^{\alpha_{\rm out}}, & f > f_p,
        \end{cases} \\
  \Omega_{\rm SBPL}(f)      &= A\,(f/f_p)^{\alpha_{\rm in}}\,
                              \left[1+(f/f_p)^{\Delta}\right]^{(\alpha_{\rm out}-\alpha_{\rm in})/\Delta}.
\end{align}
The parameters are the amplitude $A$ (quoted as $\log_{10}A$), the pivot frequency
$f_{\rm ref}$, the tilt $\alpha$, the break frequency $f_p$, the slopes $\alpha_{\rm in}$ and
$\alpha_{\rm out}$, and the smoothing width $\Delta$.  For the SBPL family $\Delta \to \infty$
recovers the BPL template and $\alpha_{\rm in}=\alpha_{\rm out}$ a plain power law.  The
$\Delta\to 0$ limit is non-standard and divergent in the usual convention.  The amplitude
parameter is related to the peak value by
$\Omega_{\rm SBPL}(f_p) = A\,2^{-(\alpha_{\rm in}-\alpha_{\rm out})/\Delta}$, and the three
families carry $3$, $4$ and $5$ parameters, the SBPL vector being
\begin{equation}
  \theta = (\log_{10}A, \alpha_{\rm in}, \alpha_{\rm out}, \log_{10}f_p, \log_{10}\Delta)\;,
\end{equation}
with all amplitude, frequency and width parameters sampled in dimensionless $\log_{10}$ form
and slopes kept linear.  Because $A$ and $f_{\rm ref}$ (or $f_p$) enter only through
$A/f_{\rm ref}^{\alpha}$, the PL model is exactly degenerate along
$\delta\log_{10}A = \alpha\,\delta\log_{10}f_{\rm ref}$.  The pivot frequency is a
parametrisation choice rather than a property of the signal, and it is fixed at
$f_{\rm ref}=1\,$mHz, so that the pair carrying the constraint is $(\ln A,\alpha)$
(Sec.~\ref{sec:calibration}).  The break frequency of the broken families is a shape
parameter, and it carries no such degeneracy.

\subsection{Detector PSDs}\label{sec:detectorpsd}
For ground-based detectors we adopt the simplified strain PSDs
\cite{Capote:2024}.  For the space-borne missions we adopt an analytic single-link Michelson
form in terms of the optical metrology system (OMS) displacement noise $P_{\rm OMS}$, the
test-mass acceleration noise $P_{\rm acc}$ and the arm length $L$,
\begin{align}
  S_h^{\rm inst}(f) &= \frac{4}{L^2}\Biggl[
      P_{\rm OMS}\Bigl(1+\bigl(\tfrac{f^{\rm OMS}_{\rm knee}}{f}\bigr)^{4}\Bigr)
      \notag\\
    &\qquad + \frac{P_{\rm acc}}{(2\pi f)^4}
      \Bigl(1+\bigl(\tfrac{f^{\rm acc}_{\rm knee}}{f}\bigr)^{2}\Bigr)
    \Biggr]\,\mathcal{R}(f)\,,
  \label{eq:spaceborne_psd}
\end{align}
where the post-time-delay-interferometry (TDI) noise-suppression factor
\begin{equation}
  \mathcal{R}(f) = \frac{\bigl[1+\tfrac{1}{2}(f/f_*)^2\bigr]}{\bigl[1+(f/f_*)^2\bigr]}\,,
\end{equation}
with the arm-transfer frequency $f_* = c/(2\pi L)$.  The factor $\mathcal{R}(f)$ multiplies the
single-link instrumental noise so that the TDI combination rejects high-frequency noise.  It is
a \emph{noise} transfer function bounded by $1\ge\mathcal{R}(f)\ge\tfrac12$, distinct from the
signal response, which is already absorbed into the sensitivity construction of
Section~\ref{sec:benchmark}.  The two high-pass knees are
$f^{\rm OMS}_{\rm knee} = 1\ {\rm mHz}$ and $f^{\rm acc}_{\rm knee} = 0.4\ {\rm mHz}$.  The
former is a simplified choice, distinct from the $2\,$mHz value of the full
Robson--Cornish--Liu 2019 model used only as the independent cross-check of
Sec.~\ref{sec:benchmark}, and is the dominant source of the factor $\lesssim3.5$ discrepancy
there, without affecting the bias ratios $R$.  For LISA we take $L = 2.5\times10^{9}\,{\rm m}$
(so that $f_* = 19.09\ {\rm mHz}$), $P_{\rm OMS}^{1/2} = 15\ {\rm pm}/\sqrt{\rm Hz}$ and
$P_{\rm acc}^{1/2} = 3\ {\rm fm\,s^{-2}}/\sqrt{\rm Hz}$.  For Taiji $L = 3\times10^{9}\,{\rm m}$,
$P_{\rm OMS}^{1/2} = 8\ {\rm pm}/\sqrt{\rm Hz}$ and the same acceleration
noise~\cite{Taiji:2017}.  TianQin is modelled by the acceleration-noise-dominated form of
\cite{TianQin:2016}.  Its arm length is $L=\sqrt{3}\times10^{5}\,{\rm km}=1.73\times10^{8}\,$m,
its geocentric orbit has radius $10^{5}\,$km and period $3.65\,$d, and it carries the same
equal-arm TDI transfer function at its own $f_*=c/(2\pi L)=0.28\,$Hz.  Its three channels are
combined in the same $A/E/T$ basis, with the TianQin--LISA and TianQin--Taiji
inter-constellation overlap reduction of Eq.~\eqref{eq:orfdef} at the corresponding
separations.  The simplified $\mathcal{R}(f)$ is
calibrated so that Eq.~\eqref{eq:spaceborne_psd} agrees with the published LISA instrumental
curve of \cite{Robson:2019} to better than twenty per cent over the $4$--$8\,$mHz span, with a
residual factor of $\lesssim3.5$ across the wider $1$--$10\,$mHz band that carries the SGWB
information.

\section{Single-parameter bias diagnostic}
\label{sec:biasdiag}

For a parameter set $\theta$ the maximum-likelihood estimator (MLE) $\hat\theta$ of
Eq.~\eqref{eq:whittle} is biased at order $1/N_s$.  For the \texttt{gauss\_D} likelihood and a
single free amplitude, Ref.~\cite{Franciolini:2025PRD} derive the per-bin MLE of the
PSD~\eqref{eq:pD}, whose fractional bias is
\begin{equation}
  \mathcal{B} \;\equiv\; \frac{\mathbb{E}[\hat P_k]-P^{\rm t}_k}{P^{\rm t}_k}
  \;=\; -\frac{1}{N_s}\,,
  \label{eq:bias_fprr}
\end{equation}
where $P^{\rm t}_k\equiv P_k(\theta_0)=S_h(\theta_0;f_k)=\mathbb{E}[\hat C_k(f_k)]$ is the
ensemble-averaged cross-PSD in bin $f_k$, and the debiasing prefactor in $\hat P^{\rm D}_k$
cancels the $\mathcal{O}(1/N_s)$ downward Student-$t$ fluctuation of $\hat C_k$ to leading
order.  The fractional statistical uncertainty per parameter is
\begin{equation}
  \mathcal{S} \;=\; \frac{1}{\sqrt{N_f^{\rm Nyq}\,N_s}}\,,
  \label{eq:stat_fprr}
\end{equation}
with $N_f^{\rm Nyq} = f_{\rm max}\,T_{\rm seg}$ the \emph{full-band} Nyquist count up to the
sampling Nyquist frequency ($f_{\rm max} = 0.5\ {\rm Hz}$ for the space-borne missions,
following Ref.~\cite{Franciolini:2025PRD}).  The resolved-bin count $N_f$ of Eq.~\eqref{eq:Nf} entering the calibration
below is a different quantity, and the two are not interchangeable.
Because $N_f^{\rm Nyq} \propto T_{\rm seg}$ and $N_s = T_{\rm obs}/T_{\rm seg}$,
\begin{equation}
  \label{eq:BSscaling}
  \frac{\mathcal{B}}{\mathcal{S}} = \sqrt{\frac{N_f^{\rm Nyq}}{N_s}}
  = T_{\rm seg}\sqrt{\frac{f_{\rm max}}{T_{\rm obs}}}
  \propto T_{\rm seg} \ ,
\end{equation}
so the bias-to-statistics ratio grows linearly with the segment length.
Table~\ref{tab:prepro} collects the diagnostic, and the
ratio of its $4\,$s and $10\,$min rows confirms the exponent of unity.  At the $11.5\,$d LISA
baseline, which averages $N_s=127$ segments over a four-year mission, the single-parameter
\emph{Gaussian} diagnostic gives $\mathcal{B}/\mathcal{S} = 62.5$.  A Gaussian-likelihood
analysis would be deeply bias-dominated.  The ground-based networks at
$T_{\rm seg}\sim4\,{\rm s}$ remain statistics-dominated, and the six-year Taiji baseline
returns a correspondingly smaller ratio because its longer mission supplies more segments.  The
inflation factor quoted throughout this work isolates the point-bias contribution to the raw
error bar,
\begin{equation}
  R_i \equiv \frac{\sqrt{(\mathbf{F}^{-1})_{ii}+b_i^2}}{\sigma_{\rm naive,i}}
        = \sqrt{1 + \left(\frac{b_i}{\sigma_{\rm naive,i}}\right)^2}\,,
  \label{eq:Rdef}
\end{equation}
with $\sigma_{\rm naive,i}=\sqrt{(\mathbf{F}^{-1})_{ii}}$ the joint marginal Fisher error.
Both $b_i$ and $\sigma_{\rm naive,i}$ are evaluated on one and the same frequency grid, the
resolved-bin analysis grid $N_f^{\rm an}$.  On that grid $R_i$ is the ratio of the calibrated
estimator's actual root-mean-square (RMS) error to the Fisher error of that same estimator.
The bar it describes is the marginal Fisher bar quoted on the resolved-bin count.  $R_i>1$ measures how much 
the uncorrected error bar is inflated by the point bias and
$R_i=1$ only once $b_i$ is removed.  The estimator covariance $\mathbf{C}_{\hat\theta}$, the
efficiency gap $\mathbf{C}_b$ and the raw-MSE matrix
$\mathbf{\Sigma}_{\rm MSE}=\mathbf{F}^{-1}+\mathbf{C}_b+\mathbf{b}\mathbf{b}^{\rm T}$ are
defined in Section~\ref{sec:vectorbias}.  The non-removable part of the raw-MSE inflation is
the efficiency gap expressed as a variance ratio,
\begin{equation}
  \rho_i \equiv \sqrt{\frac{(\mathbf{C}_{\hat\theta})_{ii}}{(\mathbf{F}^{-1})_{ii}}}
        = \sqrt{1 + \frac{(\mathbf{C}_b)_{ii}}{(\mathbf{F}^{-1})_{ii}}}\,,
  \label{eq:rhodef}
\end{equation}
so $\rho_i\ge 1$ even after the point bias is subtracted.  Its removable counterpart is $R_i$,
and a forecaster monitors the two together, since on $\alpha_{\rm out}$ the durable part
survives the removal of the point bias.  The four inflations used below are distinct.  $R_i$ is
removable by debiasing the estimator, $\rho_i$ is not, $\sqrt{1+C_b(k)}$ of
Sec.~\ref{sec:calibration} is the level-space analogue of $\rho_i$, and the parameter inflation of
Sec.~\ref{sec:params} is their projection onto the template basis.

The point-bias inflation of Eq.~\eqref{eq:Rdef} obeys a definite scaling with the
segmentation.  The point bias falls as $b\propto1/N_s\propto T_{\rm seg}$, while the
resolved count moves oppositely, $N_f\propto1/N_s$.  The Fisher error
$\sigma_{\rm naive}\propto(N_sN_f)^{-1/2}$ is therefore re-segmentation invariant at fixed
$T_{\rm obs}$ and
\begin{equation}
  R^2 - 1 \;=\; (b/\sigma_{\rm naive})^2
  \;\propto\; \frac{N_f}{N_s}
  \;\propto\; T_{\rm seg}^2\,,
  \label{eq:Rscaling}
\end{equation}
so the point-bias inflation grows quadratically with the segment length, the square of the
linear growth of the bias-to-statistics ratio $\mathcal{B}/\mathcal{S}\propto T_{\rm seg}$ of
Eq.~\eqref{eq:BSscaling}.  The efficiency gap, by contrast, is a ratio of level variances at
fixed $w_k$ and carries no $T_{\rm seg}$ scaling.  The invariance is exact.  The coarse bin
holds a fixed interval $\Delta f_k$ of a fixed observation time $T_{\rm obs}$, so re-segmenting
the data only trades the number of segments against the fine bins per coarse bin and leaves both
the per-bin level variance and the per-bin shape information as they are.  A thirty-two-fold
change in $T_{\rm seg}$ moves $1+C_b$ by less than $4\times10^{-4}$.

In $\log_{10}$-amplitude
units, Eq.~\eqref{eq:bias_fprr} together with the Jensen correction for the log-averaged
estimator gives a bias coefficient $a_{\log_{10}A} = -3/(2\ln 10) = -0.651$ per unit
$1/N_s$.  The factor $3/2$ combines the first-order bias
$\mathbb{E}[\ln\hat P]-\ln P = -1/N_s+\mathcal O(N_s^{-2})$ of the log-periodogram estimator
with the $1/(2\ln10)$ Jacobian from $\Omega\propto A^2$~\cite{Franciolini:2025PRD}.  Our
Monte-Carlo calibration reproduces this to the $1.1\%$ level, Section~\ref{sec:parammc}.

Ref.~\cite{Franciolini:2025PRD} assume both the signal and the noise to carry known
spectral shapes, so that a single amplitude describes the signal, and derives the scalar bias
of Eq.~\eqref{eq:bias_fprr} that follows.  Once the shape is free the parameter set is a
vector, so the bias becomes a vector and a bias-independent covariance term appears beside
it.  Both are obtained in closed form in the next three sections, and the template dependence
of the result is quantified in Sec.~\ref{sec:calibration}.  The multi-detector extension is
treated separately in Section~\ref{sec:network}.

\section{The efficiency gap of the two-step route}
\label{sec:gap}

\subsection{The level and nuisance block}
\label{sec:gapmodel}

A segmentation of length $T_{\rm seg}$ resolves frequency bins of width
$\Delta f=1/T_{\rm seg}$, so the physically available count in the analysed band
$[f_{\min},f_{\max}]$ is
\begin{equation}
  N_f \;=\; (f_{\max}-f_{\min})\,T_{\rm seg}\,,
  \label{eq:Nf}
\end{equation}
i.e.\ $N_f=9.93\times10^{4}$ at the $11.5\,$d baseline but only $N_f=180$ at a
$T_{\rm seg}\sim30\,$min baseline.  Let $w_i\equiv N_s$ be the statistical weight of one such
fine bin and let $X(f)=\sum_a\Omega_a(f)$ collect the signal and nuisance energy densities, so
that the share of component $a$ is $\Omega_a(f)/X(f)$.  Writing $\theta$ for the signal
parameters and $\eta$ for the nuisance direction (noise amplitudes and foreground template
coefficients), the fractional residual on fine bin $i$ is
\begin{equation}
  r_i \;=\; h_i^{\rm T}\theta \;+\; g_i^{\rm T}\eta \;+\; \epsilon_i\,,
  \qquad
  \mathrm{Var}[\epsilon_i] \;=\; \frac{1}{w_i}\,,
  \label{eq:gapmodel}
\end{equation}
with the level response $h_i \equiv (X_{\rm sig}/X_i)(1,\ln(f_i/f_{\rm ref}))$ and the
share-normalised nuisance response $g_i \equiv \Omega_a(f_i)/X_i$.  The nuisance is drawn from a
band-shared Gaussian prior $\eta\sim\mathcal N(0,\tau^2 I)$, with $\tau$ the fractional width
within which the external noise model is trusted.

Equation~\eqref{eq:gapmodel} is linear in $\phi^{\rm T}\equiv(\theta^{\rm T},\eta^{\rm T})$
and Gaussian in the noise, so it is a linear model with a design matrix and a diagonal noise
covariance,
\begin{equation}
  \mathbf r \;=\; \mathbf X\,\phi \;+\; \boldsymbol\epsilon\,,
  ~~
  \mathbf X \equiv [\,\mathbf H\ \ \mathbf G\,]\,,
  ~~
  \mathrm{Cov}[\boldsymbol\epsilon] \;=\; W^{-1}\,,
  \label{eq:linearmodel}
\end{equation}
whose rows are $h_i^{\rm T}$ in $\mathbf H$ and $g_i^{\rm T}$ in $\mathbf G$, with
$W\equiv\mathrm{diag}(w_i)$.  The weighted blocks of Eq.~\eqref{eq:gapblocks} are the level and
nuisance pieces of the information it carries.

The band is cut into $N_k$ coarse bins of $m=N_f/N_k$ fine frequencies each, and the three
sufficient objects of bin $k$ are
\begin{equation}
  w_k = \sum_{i\in k} w_i\,,
  ~~
  \beta_k = \sum_{i\in k} w_i\, g_i\,,
  ~~
  D_k = \sum_{i\in k} w_i\, g_i g_i^{\rm T}\,.
  \label{eq:gapblocks}
\end{equation}
$D_k$ retains the full within-bin resolution and therefore carries two distinct pieces.
These are the level direction, through $\beta_k$, and the residual shape,
\begin{equation}
  A_k \;\equiv\; D_k - \frac{\beta_k \beta_k^{\rm T}}{w_k}\,,
  \qquad
  c_k \;\equiv\; \frac{\beta_k}{w_k}\,,
  \label{eq:gapshape}
\end{equation}
where $c_k$ is the sensitivity of the bin level to the nuisance and $A_k$ is the part of the
bin's nuisance information that is invisible to its level.  The whole of the route dependence
below is carried by $A_k$.

\subsection{Closed form from the exact block}
\label{sec:gapclosed}

Collect the $N_k$ levels $\mu_k$ and the $p$ nuisance directions into one linear system.  With
$B\equiv[\beta_1,\dots,\beta_{N_k}]$ and $D_{\rm tot}\equiv\sum_lD_l$, the Fisher information of
$(\mu_1,\dots,\mu_{N_k},\eta)$ is the block matrix
\begin{equation}
  \mathcal I \;=\;
  \begin{pmatrix}
    \mathrm{diag}(w_k) & B \\[2pt]
    B^{\rm T} & D_{\rm tot}+\tau^{-2}I
  \end{pmatrix},
  \label{eq:gapblockmatrix}
\end{equation}
where the prior contributes $\tau^{-2}I$ to the nuisance block.  The level block of the inverse
follows from the block-inverse identity,
\begin{equation}
  \mathrm{Var}(\mu_k) \;=\; \frac{1}{w_k}
    \;+\; c_k^{\rm T}\Bigl(\sum_l A_l + \tau^{-2}I\Bigr)^{-1} c_k\,,
  \label{eq:gapclosed}
\end{equation}
because $\sum_l(D_l-\beta_l\beta_l^{\rm T}/w_l)=\sum_lA_l$ is what remains of the total nuisance
information after the level directions are projected out.  Equation~\eqref{eq:gapclosed} is the
joint route, in which the levels and the shared nuisance are determined in a single fit over the
band rather than in two stages.  Every bin contributes to the same nuisance estimate.

The two terms differ in size.  On the LISA band with the setting of
Sec.~\ref{sec:calibration}, $\mathrm{tr}\sum_lA_l$ falls short of $\mathrm{tr}\,D_{\rm tot}$ by a
factor $1.4\times10^{-3}$.  The information that the joint route exploits is therefore a small
residual of the total nuisance information, and the levels themselves supply only $1/w_k$.
Using
$D_{\rm tot}$ in place of $\sum_lA_l$ over-states the borrowed information by nearly three orders
of magnitude.

For a general template the forecast error is the mean-squared-error matrix of the estimator,
\begin{equation}
  \mathbf{\Sigma}_{\rm MSE}
    \;=\; \mathbf{C}_{\hat\theta} + \mathbf{b}\mathbf{b}^{\rm T}
    \;=\; \mathbf{F}^{-1} + \mathbf{C}_b + \mathbf{b}\mathbf{b}^{\rm T}\,,
  \label{eq:mse}
\end{equation}
with $\mathbf{b}\equiv\mathbb E[\hat\theta]-\theta$ the point bias and
$\mathbf{C}_b\equiv\mathbf{C}_{\hat\theta}-\mathbf{F}^{-1}$ the \emph{efficiency gap}, the part of
the estimator covariance that survives subtraction of the deterministic bias.
$\mathbf{F}^{-1}$ enters as a normalisation scale rather than as a Cram\'er--Rao bound, which
constrains unbiased estimators only and therefore does not bound the composite studied here.  On
the diagonal the three pieces add, so that a corrected forecast is the sum of separately
interpretable terms rather than a single rescaling.

\subsection{The first-step span as the controlling parameter}
\label{sec:gapng}

A pipeline that realises the forecast in two steps fits the shared nuisance on a limited span.
Let $n_g$ denote the number of coarse bins that share one first-step nuisance fit, the levels
remaining free inside the group.  A value $n_g=1$ is a per-bin fit and $n_g=N_k$ a single
band-wide fit.
Writing $g(k)$ for the group of bin $k$ and $A_g\equiv\sum_{l\in g(k)}A_l$ for the shape
information accumulated over it, the repeated block inversion gives the two-step level variance
and the gap
\begin{equation}
\begin{aligned}
  \mathrm{Var}_T(\mu_k) &\;=\; \frac{1}{w_k}
    + c_k^{\rm T}\bigl(A_g + \tau^{-2}I\bigr)^{-1} c_k\,,\\
  C_b(k) &\;=\; \frac{\mathrm{Var}_T(\mu_k)}{\mathrm{Var}(\mu_k)}-1\,.
\end{aligned}
  \label{eq:gapng}
\end{equation}
Equation~\eqref{eq:gapng} interpolates between the two exact endpoints.  At $n_g=N_k$ the group
is the band, the bracket coincides with that of Eq.~\eqref{eq:gapclosed}, and $C_b$ vanishes
identically.  This endpoint is used below as an internal null.  At $n_g=1$ each bin is its own
group, the bracket reduces to $A_k+\tau^{-2}I$, and the gap is maximal.  Because
$A_k$ is the only object that differs between the two ends, the gap is a direct measure of how
much within-bin nuisance shape the first step is able to use.

\subsection{Exact statements and the zero criterion}
\label{sec:gapfacts}

The gap of Eq.~\eqref{eq:gapng} is non-negative, $C_b(k)\ge0$, because the group information
of the two-step route is a sub-block of the band information and inverting a larger information
matrix never increases a marginal variance.  Over four thousand random configurations of
$(g_i,w_i,{\rm binning},\tau)$ the smallest diagonal element of $C_b$ found is consistent with
zero at the numerical precision of the computation.  It vanishes if and only if the shape
information gathered by the other bins is orthogonal to the level direction of bin $k$ in the
metric of that bin's own nuisance precision,
\begin{equation}
  c_k^{\rm T}\bigl(A_k+\tau^{-2}I\bigr)^{-1}
  \bigl(A_g - A_k\bigr)\bigl(A_k+\tau^{-2}I\bigr)^{-1}c_k \;=\; 0\,.
  \label{eq:gapzero}
\end{equation}
If the nuisance response
is constant inside a bin then $D_k=w_kg_kg_k^{\rm T}$ is rank one and $A_k=0$.  Neither
route can then constrain the nuisance from within the bin, and both fall back on the prior.
The gap vanishes for every $n_g$, and the level variance degenerates to
$1/w_k+\tau^2|g_k|^2$.  In the limit $m=1$, where each bin holds a single frequency, this is
automatic.  A bin with $A_k\ne0$ can instead be gap-free if the shapes of the
remaining bins project to zero on its level direction.  The distinction matters for
interpretation.  The gap is not a price paid for a finite nuisance prior, since a constant
response costs nothing however tight the prior.  It is the price of not sharing a
\emph{shape} constraint.  For a linear nuisance this criterion is the condition under which a
two-step estimator attains the efficiency of the joint fit~\cite{Ackerberg:2014}.

The calibration verifies both behaviours.  A configuration in which the nuisance response is
constant across the band leaves the gap at the level of $10^{-9}$, and a construction in which
bin $k$ responds only to its own nuisance direction returns machine zero.

\subsection{Scaling with the mission duration}
\label{sec:gapduration}

The gap of Eq.~\eqref{eq:gapng} is quoted at the four-year baseline.  Its dependence on the
mission duration follows from the same block.  Only the statistical weight carries the observation
time, $w_k\propto N_s\propto T_{\rm obs}$.  The level response $c_k=\beta_k/w_k$ does not, and the
shape information $A_l$ scales in proportion to it, because $D_l$ and $\beta_l\beta_l^{\rm T}/w_l$ are both
proportional to $N_s$.  Writing $s\equiv T_{\rm obs}/4\,{\rm yr}$ and keeping the four-year values
on the right-hand side,
\begin{equation}
\begin{aligned}
  \mathrm{Var}_T(\mu_k;s) &\;=\; \frac{1}{s\,w_k}
    + c_k^{\rm T}\bigl(sA_g+\tau^{-2}I\bigr)^{-1} c_k\,,\\
  \mathrm{Var}(\mu_k;s) &\;=\; \frac{1}{s\,w_k}
    + c_k^{\rm T}\Bigl(s\sum_lA_l+\tau^{-2}I\Bigr)^{-1} c_k\,,
\end{aligned}
  \label{eq:gapduration}
\end{equation}
which reduce to Eqs.~\eqref{eq:gapng} and~\eqref{eq:gapclosed} at $s=1$, the gap being the ratio
of the two.  The duration is not the segmentation of Sec.~\ref{sec:biasdiag}.  That invariance
holds at fixed $T_{\rm obs}$ and concerns $T_{\rm seg}$, while $s$ moves the segment count and the
segment length together and leaves their product fixed.

For a per-bin first step, the nuisance term of Eq.~\eqref{eq:gapduration} is a spectral average of
the nuisance precision.  In the eigenbasis of $A_k$, with eigenvalues $\lambda_i$ and weights
$p_i=(c_k\!\cdot\!v_i)^2$, it reads $V_k(s)=\sum_i p_i/(s\lambda_i+\tau^{-2})$, which approaches
the plateau $\tau^2|c_k|^2$ as $s\to0$ and decays as $s^{-1}\sum_ip_i/\lambda_i$ once the data
dominate.  Which regime a bin occupies at the baseline is decided by the projection-weighted
eigenvalue $\lambda_{\rm eff}\equiv|c_k|^2/\sum_ip_i/\lambda_i$, not by the smallest eigenvalue of
$A_k$, since the response need not project onto it.  Bin 13 has
$\lambda_{\min}=9.9\times10^{-3}$ but $\lambda_{\rm eff}=9.0$, its gap already saturated, whereas
a bin whose response stays in the slow modes still grows.  At the low-frequency end the ordering
reverses and the gap grows in proportion to the mission.

The growth rate is not fitted.  Differentiating the spectral average gives the exact identity
$-\mathrm d\ln V_k/\mathrm d\ln s=\langle \omega\rangle_V$, with
$\langle \omega\rangle_V=\sum_i\theta_i\omega_i$, $\theta_i=p_i/(s\lambda_i+\tau^{-2})/V_k$ and
$\omega_i=s\lambda_i/(s\lambda_i+\tau^{-2})$, so the slope measures the share of a bin's nuisance
information that the data carry.  On the joint arm that share is close to one for every bin and
the fit is data-limited throughout, so the logarithmic slope of the level-error ratio,
$\mathrm d\ln(1+C_b)/\mathrm d\ln s$, is the mirror image of $\langle \omega\rangle_V$.  The share
is small at the low-frequency end, where the gap grows in step with the mission, and it reaches
the joint value through the middle of the band, where the gap is already saturated.  It falls
again at the high-frequency end, where the within-bin shape degenerates and the gap resumes its
growth.  The four-year baseline sits above the saturation point of the mid-band bins and below
that of the low-frequency ones, so a longer mission buys back gap at the low-frequency end alone.

\subsection{Blindness of the gap to the periodogram likelihood}
\label{sec:gapblind}

The Gaussian per-bin estimators in use form a one-parameter family
\begin{equation}
\begin{aligned}
 && -2\ln\mathcal L_c(P|D) \;=\; N_s\,\frac{(D-P)^2}{P^2} + c\ln P\,,\\
 && \hat P_c \;\equiv\; f_c(N_s)\,D=\frac{-N_s+\sqrt{N_s^2+2cN_s}}{c}D\,,
\end{aligned}
  \label{eq:gapblind}
\end{equation}
with $\hat P_0=D$ the exact-Whittle estimator and $c=2$ the determinant-approximate Gaussian
primitive of Appendix~\ref{app:likes}.  The two estimators differ only by the multiplicative
constant $f_c(N_s)$, whose value at the baseline segmentation is $f_0=1$ and
$f_2=0.99225$.  On the logarithmic scale their variances are equal,
\begin{equation}
  \mathrm{Var}\bigl[\ln \hat P_c\bigr] \;=\; \psi'(N_s)
  \qquad\text{for all }c\,,
\end{equation}
$\psi'$ being the trigamma function.  A direct Monte-Carlo with $4\times10^5$ draws
returns the same variance for the two extreme members of the family, $c=0$ and $c=2$, to machine
precision.  Because $w_k$ is unchanged, Eqs.~\eqref{eq:gapclosed}--\eqref{eq:gapng} are unchanged
as well.  Within this family the gap belongs to the route and not to the likelihood.  The only channel through
which the likelihood of Eq.~\eqref{eq:gapblind} can move the gap is a change of the curvature
weight $w_k$.  A curvature convention re-scales that weight by $1+3/N_s$ at most, which moves
the maximum gap by a few per cent.

The practical consequence is a separation of concerns.  Replacing the working likelihood of a
forecast within the family changes the deterministic bias, because it moves the per-bin objective
relative to the sample mean, but it leaves the estimator efficiency, which is fixed by the span of
the first step.  Section~\ref{sec:calibration} quantifies the second effect and
Section~\ref{sec:parammc} verifies it against a posterior sample.

\section{In-band calibration}
\label{sec:calibration}

The reference setting for the calibration follows the published millihertz convention.  The
band is $[10^{-4},10^{-1}]\,$Hz, $N_s=127$ segments of $T_{\rm seg}=11.5\,$d, a fine grid of
$N_f=9.93\times10^{4}$ frequencies [Eq.~\eqref{eq:Nf}], $N_k=20$ coarse bins of
$m=4963$ fine frequencies each, a nuisance prior width $\tau=0.2$, and $p=2$ nuisance
directions corresponding to the test-mass acceleration and optical-metrology noise amplitudes.
The prior width is deliberately loose: $\tau$ measures how far the external noise model may be
trusted rather than how accurately the instrument is calibrated, and the inflation grows with it
(Sec.~\ref{sec:params}), so the values below sit at the conservative end of the range.  The two
nuisance directions are likewise the minimal budget, and each further direction lowers the gap, so
the tabulated inflation is an upper end of the range scanned.  The calibration is carried out on
the LISA band.  The mechanism is common to the three constellations, while the size of the effect
follows the shape content of each one's nuisance response, and the network Fisher of
Sec.~\ref{sec:network} covers Taiji and TianQin.  The reference frequency is $f_{\rm ref}=1\,$mHz
and the injected signal is a power law of index $\alpha=2/3$, the shape that carries the
quantitative results below except where a figure or table is labelled with another family.

\subsection{The gap across the band}

Table~\ref{tab:cbprofile} lists $C_b(k)$ on a subset of the twenty bins together with the factor
$\sqrt{1+C_b}$ by which the level error is inflated.  The scalar $C_b(k)$ is the excess level
variance of the two-step route over the joint route as a fraction of the joint level variance
[Eq.~\eqref{eq:gapng}], so that $\sqrt{1+C_b(k)}=\sigma_T(\mu_k)/\sigma(\mu_k)$ is the ratio of the
two-step to the joint level error.  It is the level-space counterpart of the parameter-space gap
$\mathbf{C}_b$ of Eq.~\eqref{eq:mse}.  Every inflation quoted in this work is measured against the
joint route at the same span and prior.
The gap spans more than an order of magnitude across the band (Table~\ref{tab:cbprofile}),
reaching its smallest value around five to eight millihertz and rising towards both edges, where
the nuisance response varies least across a bin.  The level-error
inflation follows the same profile,
and its spread is wide enough that a single band-averaged number would misrepresent the
reconstruction at both ends of the band.

\begin{table}[htbp]
  \centering
  \caption{Efficiency gap $C_b(k)$ on a subset of the twenty coarse bins of the LISA band
  for the power-law template, at $n_g=1$, $\tau=0.2$, $p=2$ [Eq.~\eqref{eq:gapng}].  The
  last column is the level-error inflation $\sqrt{1+C_b}$.}
  \label{tab:cbprofile}
  \small\setlength{\tabcolsep}{4.5pt}
  \begin{tabular}{crrr}
    \toprule
    bin & $f$ (mHz) & $C_b$ & $\sqrt{1+C_b}$ \\
    \midrule
    1  & 0.1194 & $1.0549\times10^{2}$ & 10.32 \\
    4  & 0.3366 & $1.2486\times10^{2}$ & 11.22 \\
    7  & 0.9487 & $2.3185\times10^{1}$ & 4.92 \\
    12 & 5.3351 & $7.3964\times10^{0}$ & 2.90 \\
    13 & 7.5360 & $5.6854\times10^{0}$ & 2.59 \\
    16 & 21.2394 & $1.8605\times10^{1}$ & 4.43 \\
    18 & 42.3781 & $5.9866\times10^{1}$ & 7.80 \\
    20 & 84.5554 & $1.3059\times10^{2}$ & 11.47 \\
    \bottomrule
  \end{tabular}
  \normalsize
\end{table}

The profile above is quoted for the power law, which is the shape carried through the
rest of the calibration.  The broken families add a break frequency, a post-break slope and a
smoothing width to the amplitude and the tilt (Sec.~\ref{sec:templates}), and the three are set
from the shape of the resolved millihertz foreground.  The break sits at $3\,$mHz, near the
turnover of the astrophysical contribution, the post-break slope is zero, so that the $f^{2/3}$
rise gives way to a flat plateau, and the smooth family is rounded with a width of one.
Carrying the same kernel through the two broken families shifts the largest per-bin factor by
less than a factor of two, while the projected parameter inflation of Table~\ref{tab:ngswitch}
at $n_g=1$ stays within about ten per cent of its power-law value.  Sweeping the break
frequency, post-break slope and smoothing width over their physical ranges spreads the per-bin
factor by at most another factor of two and moves the inflation by up to a factor of two as
well.  The differences sit at the high-frequency end of the band, where the three shapes separate
and the signal share drops.  The per-bin factor is the more template-sensitive of the two, and
the closed form and the exact null at $n_g=N_k$ hold for every family.

\begin{figure*}[tp]
  \centering
  \includegraphics[width=\textwidth]{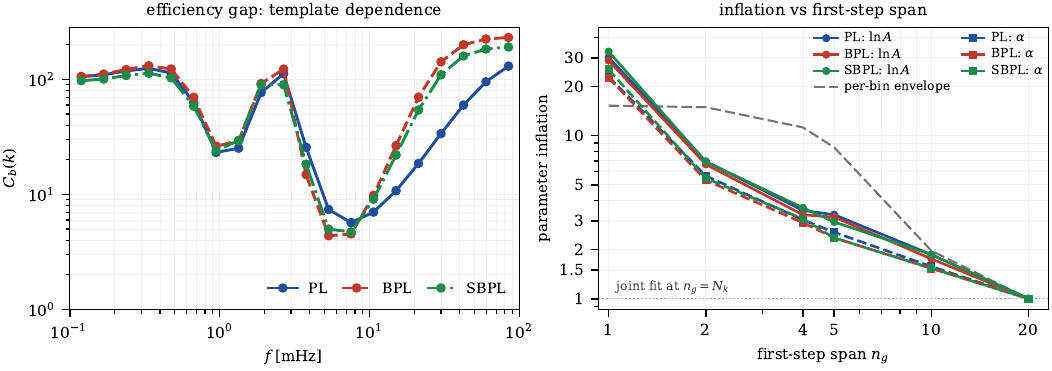}
  \caption{\textbf{Left}, efficiency gap $C_b(k)$ on the twenty coarse bins for the power law,
  the broken power law and the smooth broken power law, at $n_g=1$, $\tau=0.2$, $p=2$; the
  broken families take a break at $3\,$mHz, a flat post-break slope and, for the smooth one, a
  smoothing width of one.
  \textbf{Right}, parameter inflation against the first-step span $n_g$ for the same three
  families, on logarithmic axes, with solid curves for $\ln A$ and dashed curves for $\alpha$.
  The inflation is the ratio of the projected error on $(\ln A,\alpha)$ when the parameters are
  fitted jointly with $n_g$ bands to the error when they are fitted jointly with all $N_k$
  bands, so a single band reduces it to $\sqrt{1+C_b(k)}$, the square root of the gap drawn on
  the left.  The grey curve is the largest single-band factor of the three families.}
  \label{fig:templates}
\end{figure*}

Fig.~\ref{fig:templates} collects the calibration in both directions.  The left panel
overlays the three gap profiles on the same twenty bins.  Below about two millihertz the
three agree to within twenty per cent, the spread grows with frequency to a maximum of about
four near the upper band, and it falls back below two at the band edge.  The right
panel follows the same families across the first-step span, for the two parameters separately
and for the largest single-band factor.  Every curve falls with the span and reaches one at
$n_g=N_k$, where the first-step and the joint fits coincide, and the three families move
together, within about a tenth of the power-law curve at every span.  The largest single-band
factor lies below the projected inflation when the first step is given a single band and
above it for every span in between, with the two coinciding at the joint fit.  It therefore
overstates the inflation at every multi-band span.

Table~\ref{tab:ngswitch} is the central calibration.  It reports the gap and the parameter
inflation as the first step widens from one bin per fit to the whole band.  The inflation of the
$1\sigma$ interval on $\ln A$ and on the tilt is evaluated by projecting the level covariance
onto the template basis, $\mathbf{H}_k=(\bar\rho_k,\overline{\rho\ln f}_k)$ with $\rho_k$ the
background share of bin $k$.  This is the same projection used in
Sec.~\ref{sec:params}.  At $n_g=1$ the amplitude and tilt intervals are inflated by order thirty and
twenty.  The table shows how quickly a modest widening of the span recovers most of the loss
before the exact null at $n_g=N_k$.  The gaps at the two ends of the band fall by two to three
orders of magnitude between a per-bin and a ten-bin first step, whereas the mid-band bins are
the last to come down, being the ones that already sit closest to saturation.  The
level covariance is strongly correlated from bin to bin, so the projection uses the full
covariance rather than its diagonal, and Sec.~\ref{sec:params} records the size of the
difference.

\begin{table*}[tp]
  \centering
  \caption{First-step span $n_g$ against the efficiency gap and the projected parameter
  inflation for the power-law template, at $\tau=0.2$ and $p=2$.  The inflation is the ratio
  of the projected $1\sigma$
  interval to that of the joint route, evaluated with the full level covariance.  The last
  column is the largest off-diagonal correlation of that covariance.  The last row is exact by
  construction [Eq.~\eqref{eq:gapng}].}
  \label{tab:ngswitch}
  \small\setlength{\tabcolsep}{5pt}
  \begin{tabular}{lrrrrrr}
    \toprule
    $n_g$ & $C_b^{\min}$ & $C_b^{\rm mid}$ & $C_b^{\max}$
          & inflation $(\ln A)$ & inflation $(\alpha)$ & $\rho_{\max}$ \\
    \midrule
    $1$            & $+5.685\times10^{0}$ & $+2.567\times10^{1}$ & $+1.306\times10^{2}$ & $30.26$ & $23.01$ & $0.992$ \\
    $2$            & $+2.761\times10^{0}$ & $+8.980\times10^{0}$ & $+1.215\times10^{2}$ & $6.95$  & $5.67$  & $0.992$ \\
    $4$            & $+1.369\times10^{0}$ & $+2.755\times10^{0}$ & $+1.179\times10^{2}$ & $3.48$  & $3.05$  & $0.992$ \\
    $5$            & $+1.232\times10^{0}$ & $+2.420\times10^{0}$ & $+6.524\times10^{1}$ & $3.28$  & $2.55$  & $0.992$ \\
    $10$           & $+9.188\times10^{-2}$ & $+1.935\times10^{0}$ & $+1.935\times10^{0}$ & $1.86$ & $1.58$ & $0.992$ \\
    $20\,(=N_k)$   & $0$ & $0$ & $0$ & $1.00$ & $1.00$ & $0.992$ \\
    \bottomrule
  \end{tabular}
  \normalsize
\end{table*}

\subsection{Bin width and nuisance dimension}

The coarse-bin width and the nuisance dimension are the remaining inputs of
Eq.~\eqref{eq:gapng}, and both are scanned on the power-law shape.  Bin width sets how much
within-bin shape is available.  At fixed band
it fixes $m=N_f/N_k$, so wider bins raise $\mathrm{tr}\,A_k$.  Narrowing the reconstruction from
$N_k=5$ to $N_k=40$ lowers the maximum gap by more than an order of magnitude while leaving the
mid-band value within a factor of three of its value at $N_k=20$.  The number of nuisance
directions $p$ acts the other way.  Each
additional template direction is itself prior-limited, so the borrowed information is spread
over more directions, and the mid-band gap falls by roughly a factor of three as $p$ goes from
two to four.  Neither knob competes with $n_g$, which alone reaches the exact null at
$n_g=N_k$.  On the maximum gap the span acts at the level of two orders of magnitude, whereas
the bin width acts at about a factor of thirty and the nuisance dimension below a factor of two.

\subsection{Robustness to the foreground shape}
\label{sec:lcda}

The shapes entering the scans above are analytic stand-ins.  The first test replaces them by the
published LISA noise model \cite{Robson:2019} and the Galactic-confusion template of
App.~\ref{app:network}, and enlarges the nuisance budget from the two noise-amplitude
directions to the full template of Eq.~\eqref{eq:galc}.  A noise-only budget gives a mid-band
gap of order twenty, with an amplitude and tilt inflation of order forty to forty-seven.
Adding the Galactic amplitude and slope lowers the mid-band gap below unity, the remaining
directions moving neither by more than a few per cent.

A second test takes the foreground from an injection.  The Galactic template of
Eq.~\eqref{eq:galc} is a fit to a simulated population, and the LISA Data Challenge 2a (LDC2a)
Sangria release~\cite{LDC2a} carries a realisation of that population against which the fit can
be checked directly.  Its training file stores the Galactic binaries of the population as a
noiseless time series beside the instrumental noise and the other source families, so the
foreground spectrum is read off the injection instead of being modelled.  Over $1$--$4\,$mHz
that spectrum is nearly an order of magnitude steeper than the template of
Eq.~\eqref{eq:galc}, and in the $0.3$--$6\,$mHz anchor band it carries about $1.7$ times its
power, so the two shapes differ across the band by more than a normalisation.  For the
four-direction budget that carries the Galactic amplitude and slope, the measured shape is
rescaled to the in-band mean of the template and substituted into the kernel of
Eq.~\eqref{eq:gapng}.  This raises the mid-band $\sqrt{1+C_b}$ from $1.29$ to $2.73$ and the
largest by a similar factor, and it places the smallest gap found with the measured shape above
the largest gap found with the template.  A five-direction budget that also carries the
background amplitude absorbs the foreground normalisation into a fitted direction and brings
the two mid-band values to the same order.  Segment durations between $1$ and $11\,$d move the
mid-band value by a few per cent.  The rank criterion of Sec.~\ref{sec:gapfacts}, the two exact
limits and the monotone fall of the gap with the nuisance dimension and with the first-step
span all hold for every shape considered here.  What the shape changes is the size of the
effect at a given budget, not its dependence on the first-step span.

\subsection{Within-bin resolution}
\label{sec:ptresolution}

The calibration above retains the full within-bin resolution, and that choice is what makes the
gap finite.  If each coarse bin is represented by a single averaged frequency, $A_k$ vanishes
and the gap is identically zero for every span, at the price of discarding the shape
information that constrains the signal template in the first place.  The two limits therefore
trade against each other.  A reconstruction at the finest resolution has no route gap but also
no within-bin shape to fit a template against, whereas a coarse reconstruction has shape and
pays the gap.  The calibration reported here is at $m=4963$, where both effects are resolved,
and narrowing the bins traces the trade-off.

\section{Consequences for parameter estimation}
\label{sec:params}

\subsection{Projection onto the template basis}

The levels are not themselves the fitted quantities.  A forecast reports intervals on the
parameters of a template.  Write $\boldsymbol\mu$ for the $N_k$ levels of the band and $\mathbf C$ for their
covariance on either route, and let $\mathbf H$ be the $N_k\times2$ matrix whose $k$-th row is the
kernel $\mathbf{H}_k=(\bar\rho_k,\overline{\rho\ln f}_k)$ of Sec.~\ref{sec:calibration}.  With
$\mathbf K\equiv(\mathbf H^{\rm T}\mathbf C^{-1}\mathbf H)^{-1}$ the template parameters follow
from the generalized least-squares fit
\begin{equation}
  \hat\theta = \mathbf K\,\mathbf H^{\rm T}\mathbf C^{-1}\boldsymbol\mu\,,
  \qquad
  \mathrm{Cov}(\hat\theta) = \mathbf K\,,
  \label{eq:gls}
\end{equation}
in which the levels enter through the projector
\begin{equation}
  \boldsymbol\Pi \;\equiv\; \mathbf H\,\mathbf K\,\mathbf H^{\rm T}\mathbf C^{-1}\,,
  \qquad
  \boldsymbol\Pi^2 = \boldsymbol\Pi\,,
  \label{eq:proj}
\end{equation}
orthogonal in the $\mathbf C^{-1}$ metric and of range $\mathrm{span}(\mathbf H)$.
The joint route fits the nuisance once, so its uncertainty is set by the prior rather than by the
span of a first step, and it is the reference against which the two-step inflation is measured.
That uncertainty gives the reference interval $\sigma_{\ln A}=1.939\times10^{-3}$.  The inflation
grows monotonically with the prior width, tracking the growth of the gap and rising by nearly an
order of magnitude as $\tau$ goes from $0.05$ to $1$.  At the baseline $\tau=0.20$ it reaches
order thirty on the amplitude and order twenty on the tilt.  The projection combines the level
covariance across the band, so this parameter inflation differs from the level-error inflation
$\sqrt{1+C_b(k)}$ of Eq.~\eqref{eq:gapng}.

The convention used for the projection moves the quoted inflation by a factor of five.
Projecting the \emph{level} covariance instead of the full level-plus-template posterior costs
that factor in the joint reference interval, because compressing each bin to a single number
discards the within-bin shape that separates the template from the nuisance.  The exact block
inversion of the $(\theta,\eta)$ system agrees with the projected value to one per cent.  Using
the diagonal of the level covariance in place of the full, strongly correlated matrix
lowers the quoted inflation by the same factor of five.  All inflation ratios in this work are
therefore quoted with the full level covariance, and the absolute reference interval accompanies
them.

\subsection{Signal-only and noise-weighted limits}

Not all of the band carries signal.  Writing
$\rho_f\equiv S_h^{\rm SGWB}(f)/S_h(f)$ for the background share, the Fisher information of the
filled band is
\begin{equation}
  F_{ab} \;=\; N_s\sum_f \rho_f^2\,
     \frac{\partial\ln\Omega_{\rm GW}}{\partial\theta_a}
     \frac{\partial\ln\Omega_{\rm GW}}{\partial\theta_b}\,,
  \label{eq:fisherweighted}
\end{equation}
which reduces in the signal-only limit $\rho_f\to1$ to
\begin{equation}
  F_{ab}^{\rm (s)} \;=\; N_s \sum_f
     \frac{\partial \ln \Omega_{\rm GW}}{\partial\theta_a}
     \frac{\partial \ln \Omega_{\rm GW}}{\partial\theta_b}\,.
  \label{eq:fishersignal}
\end{equation}
A uniform suppression $\rho_f\le1$ reduces the information quadratically while changing the gap
only through the weighting of the intra-bin shape.  The signal-only limit therefore minimises
$\sigma_{\rm naive}$ and bounds the inflation conservatively at fixed gap.  The network forecast
of Sec.~\ref{sec:network} uses Eq.~\eqref{eq:fisherweighted}, with the signal-only limit retained
as the conservative bound on the inflation.

\subsection{Posterior sample}
\label{sec:parammc}

The block inverse of Sec.~\ref{sec:gapclosed} was checked against an explicit posterior sample of
the same linear model, which is the assumption of that section throughout, so the check tests the
inversion rather than the model.  The joint route is sampled with an ensemble Markov chain of $44$
walkers advanced for $4\times10^{4}$ steps each.  Discarding the first $8000$ steps and thinning
by $8$ leaves $1.76\times10^{5}$ posterior samples at an acceptance rate of $0.28$.  The two-step
route is sampled hierarchically, drawing the first-step nuisance from its own group posterior and
then the levels conditionally, from $2\times10^{5}$ draws per span.  Both samples are projected
onto $(\ln A,\alpha)$ with the kernel of Eq.~\eqref{eq:gls}.

\begin{figure}[htbp]
  \centering
  \includegraphics[width=\linewidth]{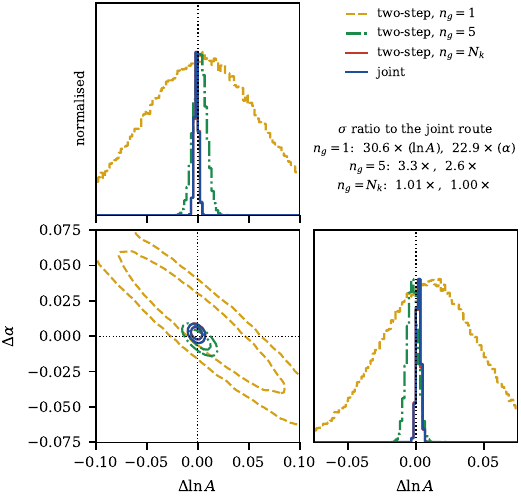}
  \caption{Triangular posterior of the power-law template parameters $(\ln A,\alpha)$ from the
  Monte-Carlo sample, drawn as the offsets $\Delta\ln A=\ln A-\ln A_{\rm inj}$ and
  $\Delta\alpha=\alpha-\alpha_{\rm inj}$ from the injected template.  The two-step
  route is drawn at the first-step spans $n_g=1$, $5$ and $N_k$, against the joint route.  The
  posterior narrows monotonically with the span and meets the joint route at the band-wide
  span.  At $n_g=1$ it is wider by a factor ${\simeq}30$ in $\ln A$ and ${\simeq}22$ in the
  tilt, as quoted in the panel.}
  \label{fig:mccorner}
\end{figure}

The sample reproduces the closed joint and two-step intervals within the sampling error of the
draws, a few tenths of a per cent at $2\times10^{5}$ samples.  It returns an amplitude inflation of
$30.6$ and a tilt inflation of $22.9$ at $n_g=1$, against closed-form values of $30.26$ and
$23.01$, and unity on the band-wide null.
Fig.~\ref{fig:mccorner} shows the resulting triangular posteriors of $(\ln A,\alpha)$, drawn
as offsets from the injected template, for the two routes at a centred prior.  The two-step
triangle is wider than the joint triangle by the
same inflation at every level, and widening the first-step span towards the band-wide fit
closes the difference.  The displacement of the two triangles when the prior is misplaced is
the bias channel of Sec.~\ref{sec:vectorbias}.  Placing the injected nuisance at half the
prior width, rather than at its centre, moves the returned ratios by at most a few per cent,
comparable to the sampling error, and translates the two triangles together rather than
apart.

\section{Consequences for signal reconstruction}
\label{sec:recon}

A binned reconstruction does not fit a template across the band.  The fitted objects are the
levels themselves, so the route enters directly.  The fractional level error of the two-step
route exceeds that of the joint route by a factor of a few to above ten across the band.  It
follows the gap profile and is largest at the two edges, where the nuisance response is least
structured.  The band-wide span returns the joint values, again confirming that the gap and not
the binning is responsible.

The factors above are per-bin ones.  The joint level covariance matrix, with its diagonal $1/w_k$
removed, is $\mathcal C\mathbf{G}\mathcal C^{\rm T}$, where $\mathcal C$ collects the per-bin
shape vectors $c_k$ and $\mathbf{G}=(\sum_lA_l+\tau^{-2}I)^{-1}$.  That matrix has rank $p$, the
nuisance dimension, so the joint route recovers information only along the subspace that
$\mathcal C$ spans, $p=2$ here, and directions orthogonal to it gain nothing.  The amplitude
direction, a band average that resolves no structure within the band, is improved by $2.3$.  A
reconstruction that does resolve such structure is penalised in the worst direction, which has
the closed form
$\sqrt{1+w_k\,\lambda_{\max}(\mathbf{P}\,\mathrm{diag}(\mathrm{Var}_T)\mathbf{P})}=125.3$, with
$\mathbf{P}\equiv I-\mathcal C(\mathcal C^{\rm T}\mathcal C)^{-1}\mathcal C^{\rm T}$
the orthogonal projector onto the complement of $\mathrm{span}(\mathcal C)$ in $\mathbb R^{N_k}$.
The level-error inflation
quoted at the start of this section lies between these two limits.

Averaging over $M$ neighbouring bins changes the picture.  Both level errors fall with $M$, as
the statistical term averages down as $1/\sqrt M$.  The nuisance channel is common across the
band and is diluted only by the overlap of the averaged responses.  The ratio of the common-mode
to the statistical contribution therefore grows monotonically with $M$, by nearly an order of
magnitude between $M=2$ and $M=20$.  It tracks $\sqrt M$ times the ratio of the level-weighted
mean nuisance response of the averaged bins to the per-bin response.  Band-averaging thus suppresses the term that would
have averaged away and leaves the term that does not, acting on the variance rather than on the
mean.

\begin{figure}[htbp]
  \centering
  \includegraphics[width=\linewidth]{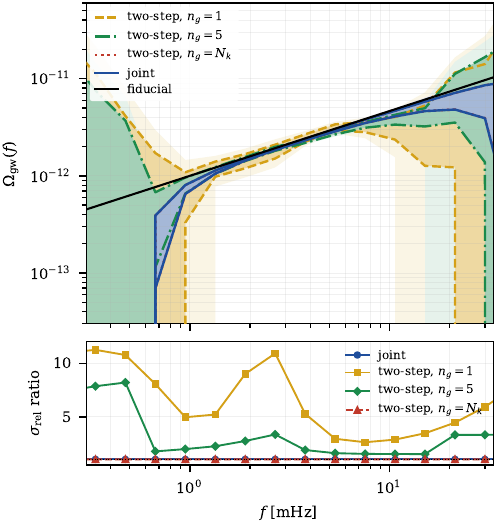}
  \caption{\textbf{Top}, reconstructed posterior band of $\Omega_{\rm GW}(f)$ for the joint
  route and the two-step route at the first-step spans $n_g=1$, $5$ and $N_k$, against the
  injected power-law spectrum (black line).  Shading spans the $95\%$ interval about a darker
  core at
  $68\%$, whose edges carry the colours and line styles used in the ratio panel.
  \textbf{Bottom}, ratio of the per-bin fractional errors to the joint route.  Every two-step
  band is wider at every frequency, and the band-wide span approaches the joint route from
  above.}
  \label{fig:mcrecon}
\end{figure}

Fig.~\ref{fig:mcrecon} shows the reconstructed posterior band of the two routes over the
$0.3$--$30\,$mHz window that carries the SGWB information, together with the ratio of the
per-bin errors.  The two-step bands are wider at every frequency, and the joint band coincides
with the band-wide first-step span.

\section{The bias-vector channel}
\label{sec:vectorbias}

The gap is the bias-independent part of the estimator error.  The complementary channel is the
deterministic shift produced when the nuisance sits away from the centre of its prior, as it
does whenever the instrument model is mis-specified.  For a prior misplaced by $\eta_0$, the
first-step posterior contracts to $\mathbb E[\hat\eta]=(A_g+\tau^{-2}I)^{-1}A_g\eta_0$ with $A_g$
the group shape information, and the induced level bias is
\begin{equation}
  \delta_k \;=\; c_k^{\rm T}\Bigl(A_g+\tau^{-2}I\Bigr)^{-1}\tau^{-2}\eta_0\,,
  \label{eq:biasvector}
\end{equation}
so that the two routes share a single response kernel and differ only in the group over which
$A_g$ is accumulated.  In the limit in which the group carries no shape information the
expression reduces to $\delta_k=g_k^{\rm T}\eta_0$, the full shift of the level.  The ratio
$\delta_k/(g_k^{\rm T}\eta_0)$ therefore measures how faithfully a bin transmits the nuisance
misplacement, and it is unity only for a bin with $A_k=0$.

The gap shares this kernel.  The rank-two correction to the joint level covariance is built from
the same per-bin shape vectors $c_k$ that transmit the bias here, so the bias and the gap are two
readings of one object: a per-bin shift carried through $c_k$, and a band-wide information
deficit carried by the subspace those vectors span.

The level bias for a ten-per-cent prior misplacement follows the same profile as the gap because
it is driven by the same response kernel.  On the two-step route it reaches a tenth in fractional
terms at the band edges and a few per cent in the mid-band, while the joint route suppresses it by
two to three orders of magnitude.  The transmission ratio $r_k$ of Eq.~\eqref{eq:biasvector} is
close to unity at the low-frequency edge and falls to about a tenth around $20\,$mHz, where the
within-bin shape is best resolved.  It recovers part of the way towards the high-frequency edge.
A bin therefore carries the misplacement to its level almost unattenuated when its response has
little shape, and damps it when the shape is present.  Projected onto the template basis the two-step route
shifts $\ln A$ by about two-thirds of its own inflated interval, while the joint route leaves a
shift smaller by five orders of magnitude.  A band average does not remove the two-step shift,
which is common across the band and there remains comparable to the statistical error of the
averaged level.

\section{Network Fisher and Galactic confusion foreground}
\label{sec:network}

The Whittle Fisher extends to a network through the exact multi-channel
spectral-covariance inversion of Eq.~\eqref{eq:networkfisher}~\cite{Smith:2019}, with the
channel response $\mathcal R_A=\mathcal R_E=1$, $\mathcal R_T=0$ in the A/E/T basis and the
overlap-reduction functions (ORFs) at their physical (signed) values.  In the single-detector limit Eq.~\eqref{eq:networkcov} gives
$C_{ii}=S_{h,i}+\Gamma_{ii}S_h^{\rm SGWB}$, so the signal-to-total ratio entering the Fisher is
$\Gamma_{ii}\rho_f$ and the channel Fisher is $\Gamma_{ii}^2$ times the ORF-free Fisher of
Eq.~\eqref{eq:fisherweighted}, with the same bare cosmological $S_h^{\rm SGWB}(f)$ in both.
With the long-wavelength normalisation $\Gamma_{AA}(0)=\Gamma_{EE}(0)=1/2$ that
Eq.~\eqref{eq:orfdef} fixes, the A and E channels sit a factor $1/4$ in $F$ ($1/2$ in $\sigma$)
below that ORF-free form, and $\mathcal R_T=0$ removes the $T$ channel from the signal.  No
quoted number depends on the ORF normalisation.  Every forecast here is a ratio of error bars on
one common network Fisher with one common set of overlap reductions, so the factor cancels.  The
ORF-free normalisation enters only where a single-channel error bar is compared with a published
one.  Appendix~\ref{app:network} derives the adopted ORF forms, the Galactic foreground treatment
and the $T$-channel leakage.

Running the multi-channel LISA/Taiji/TianQin network Fisher at the fiducial 11.5-d
segmentation gives the relative $\sigma$ gains of Table~\ref{tab:networkforecast}, including a
headline Taiji/LISA gain of $1.25$ on $\sigma_{\log_{10}A}$ that is dominated by Taiji's lower
optical-metrology noise.  The inter-constellation overlap reduction is evaluated with the
separation argument $x=2\pi fd/c$ of Eq.~\eqref{eq:orfdef} (App.~\ref{app:network}).  Because
the exact separated-detector ORF is instead an oscillatory spherical-Bessel expansion, we check
the convention dependence directly.  The plane angle, a zero inter-constellation correlation,
the arm-transfer scale and the standard spherical-Bessel form normalised to the same co-located
limit all move the four-channel $\sigma_{\log_{10}A}$ ratio by five per cent of its adopted
value.  Most of the four-channel gain
comes from combining two sets of statistically independent A/E channels rather than from the
cross-correlation, so it is set by the Taiji-only row and is independent of this convention.  The
TianQin row is marginal.  With LISA and Taiji already combined, adding TianQin leaves both
$\sigma$ ratios unchanged, because the inter-constellation overlap with the other two
constellations is strongly suppressed over the signal band.
\begin{equation}
  F^{\rm net}_{ab} = N_s\sum_f
    \mathrm{Tr}\!\Bigl[
      C(f)^{-1}\,\partial_a C(f)\,
      C(f)^{-1}\,\partial_b C(f)
    \Bigr],
  \label{eq:networkfisher}
\end{equation}
where
\begin{equation}
  C_{ij}(f) = \delta_{ij}\,S_{h,i}^{\rm tot}(f)+ \bigl(\mathcal R_i\mathcal R_j\,\Gamma_{ij}(f)\bigr)\,S_h^{\rm SGWB}(f),
  \label{eq:networkcov}
\end{equation}
with $\partial_a C_{ij}(f)=\bigl(\mathcal R_i\mathcal R_j\,\Gamma_{ij}(f)\bigr)\,
\partial_a S_h^{\rm SGWB}(f)$, and the ORF $\Gamma_{ij}$ entering with
its signed value.  The intra-constellation off-diagonal terms vanish identically
for an isotropic background (Sec.~\ref{sec:orf}), while the inter-constellation
correlations are retained with their true sign.

The per-pair approximation of much of the earlier literature follows from keeping only the
$O\!\bigl([S_h^{\rm SGWB}]^2\bigr)$ term, which collapses the trace to a sum over detector
pairs,
\begin{align}
  F^{(\rm pair)}_{ab}
    &= N_s \sum_f \sum_{i,j}
       \frac{\bigl(\mathcal R_i\mathcal R_j\,\Gamma_{ij}(f)\bigr)^2\,
             \bigl[S_h^{\rm SGWB}(f)\bigr]^2}
            {C_{ii}(f)\,C_{jj}(f)}
       \notag\\
    &\qquad \times \partial_a\ln\Omega_{\rm GW}\,
              \partial_b\ln\Omega_{\rm GW}.
  \label{eq:networkfisher_pair}
\end{align}
Here the overlap reduction enters only as $\Gamma_{ij}^2$, so the sign of the surviving
inter-constellation off-diagonal ORFs is lost, whereas the exact inversion of
Eq.~\eqref{eq:networkfisher} keeps it at the next order, where the signed cross-ORF appears
linearly.  With $C_{ii}=S_{h,i}+\mathcal R_i^2\Gamma_{ii}S_h^{\rm SGWB}$
[Eq.~\eqref{eq:networkcov}] the per-pair sum is identical to the exact Fisher whenever
$C$ is diagonal.  On the same grid the only difference is therefore the discarded
inter-constellation structure, worth $0.80\%$ on $\sigma_{\log_{10}A}$ and $0.63\%$ on
$\sigma_\alpha$ for LISA+Taiji and nothing for LISA and Taiji alone.  The exact Fisher is used
throughout.  Every
comparison drawn here is a ratio of error bars on one common network Fisher with one common set
of overlap reductions, so a common rescaling $\boldsymbol\Gamma\to\kappa\,\boldsymbol\Gamma$
cancels.  The inflation $R$ is $\kappa$-invariant only in the strong-signal limit $\rho_f\simeq1$,
where $\boldsymbol\Gamma$ drops out of Eq.~\eqref{eq:fishersignal}.  In the noise-weighted
regime $\sigma$ scales as $1/\sqrt{\kappa}$ while the point bias $\mathbf{b}$ does not, so the
numerics quoted here are those of the physical detectors of Section~\ref{sec:detectorpsd}.

\begin{table*}[tp]
\small
  \centering
  \caption{Relative $1\sigma$ gains on $\log_{10}A$ and $\alpha$ for a power-law
  background, over the LISA, Taiji, LISA+Taiji and LISA+Taiji+TianQin networks, normalised
  to the LISA-only
  two-channel values, with the Galactic confusion foreground treated as effective
  noise.  The ratios are independent of the segmentation because every
  configuration scales as $\sigma\propto1/\sqrt{N_s}$.}
  \label{tab:networkforecast}
  \begin{tabular}{lcc}
    \toprule
    Network & $\sigma_{\log_{10}A}/\sigma^{\rm LISA-only}_{\log_{10}A}$ &
             $\sigma_\alpha/\sigma^{\rm LISA-only}_\alpha$ \\
    \midrule
    LISA only (2 ch)                    & 1.00  & 1.00 \\
    Taiji only (2 ch)                   & 0.80  & 0.61 \\
    LISA + Taiji (4 ch)                 & 0.62  & 0.51 \\
    LISA + Taiji + TianQin (6 ch)       & 0.62  & 0.51 \\
    \bottomrule
  \end{tabular}
\end{table*}

\section{Cross-checks on synthetic LISA data}
\label{sec:benchmark}

\begin{figure*}[tp]
  \centering
  \includegraphics[width=0.90\linewidth]{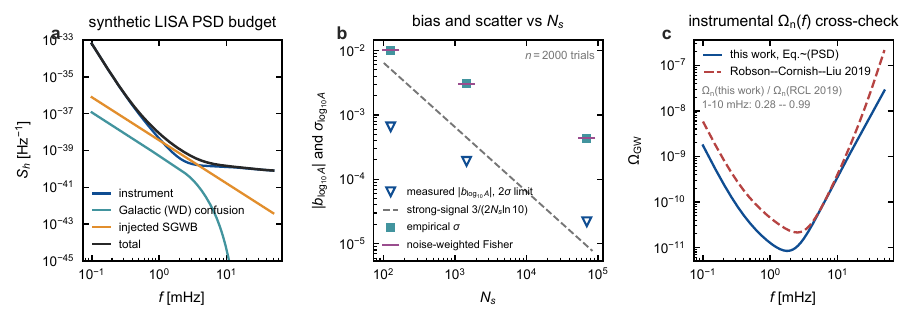}
    \caption{Cross-checks on synthetic LISA data.  \textbf{a}, synthetic PSD budget,
  instrumental noise (blue), Galactic confusion foreground (teal), an injected power-law SGWB
  $\Omega_{\rm GW}=10^{-11}(f/{\rm mHz})^{2/3}$ (orange) and their sum (black).  \textbf{b},
  amplitude bias and scatter versus $N_s$.  Blue downward triangles, $2\sigma$ upper limits;
  teal squares, empirical scatter; violet, the noise-weighted Fisher of
  Eq.~\eqref{eq:fisherweighted}; grey dashed, the strong-signal $1/N_s$ law $3/(2N_s\ln10)$.
  \textbf{c}, instrumental noise floor $\Omega_{\rm n}(f)$ from
  Eq.~\eqref{eq:spaceborne_psd} (blue) and from the Robson--Cornish--Liu 2019 expression (red
  dashed).}
  \label{fig:bench}
\end{figure*}

We cross-validate the implementation against the published sensitivity curve, against the bias
law under instrumental noise, and end to end at high signal strength.  Fig.~\ref{fig:bench}
collects the three.

Fig.~\ref{fig:bench}a shows the synthetic
PSD budget.  It carries instrumental noise, Galactic confusion foreground, and an injected SGWB
$\Omega_{\rm GW} = 10^{-11}(f/{\rm mHz})^{2/3}$ at the LISA sensitivity floor.  The
null test of Fig.~\ref{fig:bench}b draws periodogram ensembles from that total PSD at the
$11.5\,$d, $24\,$h and $30\,$min segmentations,
spanning more than two decades in $N_s$ over a four-year mission, on a $400$-bin grid over
$0.1$--$50\,$mHz.  They are fitted with the joint exact-Whittle maximum-likelihood estimator of
Eq.~\eqref{eq:whittle}, with the noise PSD held fixed, on two thousand trials at each
segmentation.  The measured bias is consistent with zero at all three segmentations, so the
point-bias inflation is bounded at the level of $10^{-3}$ on this grid; the $R^2-1\propto N_f$
scaling of Eq.~\eqref{eq:Rscaling} turns that bound into $R\le1.12$ on the physical resolved
count, and the empirical scatter reproduces the noise-weighted Fisher of
Eq.~\eqref{eq:fisherweighted} to about one per cent.  Because the injected signal sits at the
floor ($\rho_f\lesssim1$) this null test covers the weak-signal regime only; its $2\sigma$ limits
already lie well below the strong-signal $1/N_s$ law, which is the object of
Section~\ref{sec:biasdiag}.

Fig.~\ref{fig:bench}c compares the instrumental noise floor
$\Omega_{\rm n}(f) = (4\pi^2 f^3/3H_0^2)\,S_h^{\rm inst}(f)$ implied by the PSD of
Eq.~\eqref{eq:spaceborne_psd} with the published Robson--Cornish--Liu 2019
construction~\cite{Robson:2019}.  The two curves agree to better than
$20\%$ over the $4$--$8\,$mHz span, and the analytic curve stays within a factor of
$\lesssim3.5$ of the published one over the wider $1$--$10\,$mHz span that carries the SGWB
information.  This accuracy of a simplified
analytic PSD suffices for the strong-signal, ratio-based
conclusions of this work, which are governed by how far the background is resolved above the
noise, $\rho_f$, rather than by the absolute noise level (Sec.~\ref{sec:params}).  The
weak-signal entries, where the background sits near the floor, are correspondingly more
sensitive to the noise normalisation.

Since the calibration
itself is a signal-only statement, a forecast needs the calibrated inflation under realistic
noise.  We repeat that test on the same full PSD on an $8000$-bin analysis grid over the same band
with $N_s=127$, raising the injected amplitude in six steps of two thousand trials each.  Those
steps carry the background from a sub-dominant share of the total PSD to a dominant one.  The
marginal noise-weighted Fisher error on $\log_{10}A$ tracks the measured ensemble scatter of the
fitted amplitude to a few per cent at every loudness, across the whole regime and not at a single
point, which is what Eq.~\eqref{eq:fisherweighted} is meant to supply.  The composite arm behaves
differently.  It carries no nuisance parameter and no first-step span, so it is an implementation
check rather than a measurement of the route.  Its raw error bar grows monotonically with the
resolved fraction, from order five at the sensitivity floor to an order of magnitude at
saturation, and approaches the signal-only value there, while the exact-Whittle arm returns unity
throughout.  The suppression at lower loudness is a property of the noise weighting, not of the
estimator.  The convention of the log-space outer objective enters the recovered bias only at
second order, quantified in App.~\ref{app:twostep}.  In the
signal-only limit the composite estimate recovers the analytic amplitude bias to about one
per cent, which checks the amplitude channel alone.  For the smooth power law the two-step
amplitude bias coincides with the single-parameter coefficient of
Eq.~\eqref{eq:bias_fprr}.  That coincidence carries no information about the multi-parameter
inflation, which has no single-parameter analogue.

\section{Discussion}
\label{sec:discussion}

\subsection{Where the operational forecasts sit}
\label{sec:opaxis}
The binned millihertz reconstruction of Refs.~\cite{Flauger:2021,Kume:2025} splits the band
adaptively and fits each bin with a signal template of assumed shape plus a noise
model~\cite{Caprini:2019sgwb}, working with a Gaussian objective and, where the non-Gaussianity of
the unaveraged data matters, its log-normal replacement~\cite{Hamimeche:2008}.  It and the
ground-based cross-correlation
search~\cite{Matas:2021} settle the likelihood side of the question, the latter by the
approximate sufficiency of its binned statistic.  Those choices act on the working objective,
which Sec.~\ref{sec:gapblind} shows to leave the estimator efficiency untouched for the Gaussian
family it covers, on which the joint route keeps its advantage even for a working objective that
departs from the true likelihood, under a finite set of testable moment conditions~\cite{Kim:2025}.

The quantity that acts on the other axis is the span of the first-step nuisance fit.  The
published recipes put the noise model ahead of the signal.  The joint characterisations of the
instrumental noise and of the stochastic backgrounds assume the spectral shape of both to be
known, with parametrised templates supplying it~\cite{Santini:2025}.  A LISA search without a
quiet stretch is described in the review of Ref.~\cite{Renzini:2022} as one that fixes a noise
model and iterates over the data until the signal converges to the maximum-likelihood estimate,
and a cross-correlation search in the low-signal limit likewise estimates its noise covariance
independently of the signal.  The global fit of Ref.~\cite{Katz:2024} converges its PSD and
foreground sampling before searching for signals, and reports a foreground above the level
expected of the population it models.  That excess is consistent with a foreground estimated
before the signal is subtracted, an effect the same work leaves unquantified.

Where these recipes sit on the span axis is not something their descriptions settle, and it is
the number that decides the outcome.  A nuisance fitted once over the band, as in a template
search that estimates a single noise model, sits at the band-wide end, where the gap vanishes.  A
binned reconstruction that fits each level with its own noise model sits at the per-bin end of
that axis.  A noise model taken from instrument characterisation and then held fixed has no
in-band first step at all, and its level error sits at the plateau $\tau^2|c_k|^2$ of
Eq.~\eqref{eq:gapduration}, which the per-bin entry of Table~\ref{tab:cbprofile} falls below
because the bin still contributes its own shape.  The tabulated span therefore brackets the range,
and $n_g=1$ is a representative rather than an extreme.

\subsection{What a forecast should report}
\label{sec:actionable}

The bias vector and the efficiency gap of Secs.~\ref{sec:gap} and~\ref{sec:vectorbias} are the
two quantities that carry a two-step forecast beyond its Fisher scale.  The deliverable is the
measured estimator covariance together with the bias-inflated mean-squared-error matrix,
$\mathbf{C}_{\hat\theta}$ and
$\mathbf{\Sigma}_{\rm MSE}=\mathbf{C}_{\hat\theta}+\mathbf{b}\mathbf{b}^{\rm T}$, both positive
semi-definite by construction and requiring no clipping.  The debiased interval is read off
$\mathbf{C}_{\hat\theta}$ after subtracting $\mathbf{b}$, and the raw interval off
$\mathbf{\Sigma}_{\rm MSE}$.  The span of the first-step nuisance fit decides the outcome.  Fitting
the shared nuisance once per bin and fitting it once per band differ by more than an order of
magnitude in the level error, and by a factor $30$ in the amplitude interval at the baseline prior
width, while both are described by the same likelihood.  The gap is quoted at the four-year
baseline, and a longer mission moves it along the scaling of Sec.~\ref{sec:gapduration}.  The pair
(gap profile, projected interval) of Tables~\ref{tab:cbprofile} and~\ref{tab:ngswitch} is therefore
the object to carry.  It is indexed by the first-step span and by the projection-weighted
eigenvalue $\tau^2\lambda_{\rm eff}$ of Sec.~\ref{sec:gapduration}, which fixes whether a bin is
data-limited or prior-limited at the baseline.

The methodology, and every number quoted here, is reproducible from the Python package
\texttt{gwforecast} deposited with this manuscript.  It provides the likelihood primitives of
Table~\ref{tab:likes}, the segment recommender, the detector PSDs, the closed-form gap and bias
kernels of Secs.~\ref{sec:gapclosed} and~\ref{sec:vectorbias}, and the network Fisher.  It also
provides the scripts that regenerate every figure and table, the tabulated products on which the
quantitative claims rest (among them the gap profile and span scan of
Tables~\ref{tab:cbprofile} and~\ref{tab:ngswitch}), a test suite and a user guide.  All
Monte-Carlo draws use NumPy's PCG64 generator (\texttt{numpy.random.default\_rng}) with the
deterministic base seed $20261009$, so every number is reproducible without a separate
configuration file.

\subsection{Limitations}
\label{sec:limits}

The gap and the bias vector are derived for a band-shared nuisance with a Gaussian prior, and
the transferred quantities are ratios, which are insensitive to the overall normalisation of the
response but depend on the within-bin shape.  The calibration therefore states what the route
costs at a given shape content, and a treatment with a richer nuisance basis in each bin would
reduce the gap at fixed span without changing the mechanism.  Restoring the within-bin response
to a full time-delay interferometry transfer function and a template foreground, rather than the
two-direction response used here, would move the absolute gap through the nuisance dimension $p$
while leaving the span dependence intact.  The foreground is held fixed rather than jointly
fitted, its shape taken from the LDC2a injection, so the anchor of Sec.~\ref{sec:lcda}
applies to the total Galactic-binary power of the release rather than to the residual of a
global fit~\cite{Katz:2024}.  The test reads the released population spectrum directly, so it
isolates the shape content of that population and is not a data-driven foreground estimate;
testing the latter would require the foreground to be fitted from the same data as the signal.

Noise is taken stationary per segment, and glitches and gaps reduce the effective segment count
with a duty-cycle uncertainty of order ten per cent.  The deterministic bias scales as $1/N_s$,
so this rescales that channel by the inverse usable-segment fraction, and it does not enter the
gap, which is a ratio of level variances at fixed $w_k$.  A misspecification of order $N_s^0$,
such as a wrong template shape or non-stationary residuals not absorbed into the PSD, does not
vanish under finer segmentation~\cite{White:1982}, so on real data the route correction removes
the $1/N_s$ deficit while a full Bayesian or exactly specified treatment remains the appropriate
baseline wherever the $N_s^0$ term dominates.  The analytic PSD of Eq.~\eqref{eq:spaceborne_psd}
matches the published Robson--Cornish--Liu curve to better than twenty per cent over the
$4$--$8\,$mHz span (Fig.~\ref{fig:bench}c), and the Galactic
confusion foreground enters as effective noise rather than being jointly
fit~\cite{Pieroni:2020}.  The treatment is first-generation TDI with the equal-arm
approximation, second-generation TDI~\cite{Wang:2019} suppressing the residual mismatch further
(App.~\ref{sec:tchannel}).  Neither affects the ratios, while the absolute sensitivities carry
the systematic.  Under a Bayesian reading the gap and the bias survive as corrections to the
sampling distribution, and marginalisation over the nuisance does not by itself remove a
leading-order sampling bias.  A mis-specified likelihood yields a biased posterior as well as a
biased estimator, so both terms must be budgeted alongside the $N_s^0$ term.

\section{Conclusions}
\label{sec:conclusions}

Forecasts of stochastic-background parameters for the millihertz observatories derive their
error bars from a Whittle likelihood whose finite-segmentation bias has not been propagated to
the reported intervals.  We have separated the error of a two-step forecast into a bias vector
$\mathbf{b}$ and an efficiency gap $\mathbf{C}_b$, the two quantities that carry it beyond its
Fisher scale, and have calibrated both in closed form.

For a nuisance shared across the band, the information that a joint fit borrows from the other
bins is carried by the within-bin shape $A_k=D_k-\beta_k\beta_k^{\rm T}/w_k$ of the nuisance response.
The efficiency gap of a two-step realisation follows from the block inverse of the
level-and-nuisance system, Eq.~\eqref{eq:gapng}.  Both endpoints are exact, the gap vanishes
identically when the first step spans the band, and it vanishes whenever the response carries no
within-bin structure, $A_k=0$, which is not a statement about the width of the
prior.  Across the LISA band at the published segmentation and prior the gap spans more than an order
of magnitude (Table~\ref{tab:cbprofile}), and the projected $1\sigma$ inflation of the amplitude
and tilt falls from order thirty and twenty for a per-bin first step to unity for a band-wide one
(Table~\ref{tab:ngswitch}).  On the maximum gap, the bin width and the nuisance dimension act
through the same kernel at the level of a factor of thirty at most, whereas the span alone drives
the gap to an exact null at $n_g=N_k$.  In reconstruction
coordinates the per-bin level error is inflated by a factor of a few to above ten
(Table~\ref{tab:cbprofile}), the band-averaged amplitude by about two and a direction that
isolates shape by a further order of magnitude.  Band-averaging
dilutes only the statistical term.

The choice of periodogram likelihood within the Gaussian family does not change the efficiency,
because the per-bin estimators in use have the same logarithmic level variance $\psi'(N_s)$
whichever member is adopted.  The family therefore moves the deterministic bias, through the
position of the per-bin objective relative to the sample mean, and leaves the gap untouched.
Changing the working likelihood of a forecast and changing the span of its first step are
independent operations.  A posterior sample of the same model reproduces the closed form within
its sampling error on both routes, returns unity on the band-wide null, and supplies triangular
posteriors of the fitted parameters.  A ten-per-cent prior misplacement shifts the two-step levels
by up to $10^{-1}$ and the joint levels by $10^{-4}$, so the bias channel and the efficiency
channel respond to the same route parameter.

A benchmark on the LDC2a injection replaces the Galactic template by the spectrum of the
released population and moves the mid-band gap by an order of magnitude, the smallest gap found
with the measured shape exceeding the largest found with the template, so the foreground shape
joins the first-step span as an input a forecast states alongside its intervals.  The
operational consequence is a reporting requirement rather than a correction to any particular
published number.  A forecast that realises its likelihood in two steps should quote the span of
its first-step nuisance fit together with its intervals, since that number decides whether the
intervals are the Fisher scale or an order of magnitude above it.  No choice of periodogram
likelihood within the family can compensate for it.

\begin{acknowledgments}
BQL is supported in part by the National Natural Science Foundation of China under Grant No. 12405058. 
\end{acknowledgments}

\appendix
\section{Likelihood formulae}
\label{app:likes}

Table~\ref{tab:likes} lists the likelihood primitives used in this work.  All are single-bin,
single-detector forms that treat the periodogram as Gaussian (Whittle, \texttt{gauss\_S},
\texttt{gauss\_Q}, \texttt{gauss\_D}).  The forecasts of this work use the exact Whittle
likelihood of Eq.~\eqref{eq:whittle}, the determinant-approximate \texttt{gauss\_D} likelihood
appearing as the contrast arm; the single-bin forms of both are

\begin{table*}[tp]
  \centering
  \small
  \caption{Likelihood primitives, their single-bin single-detector objective, the
  corresponding per-bin maximum-likelihood estimator and its fractional bias
  $\mathcal{B}=\langle\hat P_k\rangle/P_k-1$; only \texttt{gauss\_D} retains the
  log-determinant normalisation and is therefore biased.}
  \label{tab:likes}
  \begin{tabular}{@{}p{0.20\linewidth}p{0.10\linewidth}p{0.35\linewidth}p{0.08\linewidth}@{}}
    \toprule
    Primitive & $-2\ln\mathcal{L}$ & Per-bin MLE & $\mathcal{B}$ \\
    \midrule
    \texttt{whittle} (exact) & Eq.~\eqref{eq:lW} & $\hat P^{\rm W}_k=D_k$ & $0$ \\
    \texttt{gauss\_S} & Eq.~\eqref{eq:lS} & $D_k$ & $0$ \\
    \texttt{gauss\_Q} & Eq.~\eqref{eq:lQ} & $D_k$ & $0$ \\
    \texttt{gauss\_D} & Eq.~\eqref{eq:lD} & $D_k\tfrac{N_s}{2}\bigl(\sqrt{1+4/N_s}-1\bigr)$ & $-1/N_s$ \\
    \bottomrule
  \end{tabular}
  \par\vspace{3pt}\footnotesize\emph{Note.} The third-column estimators are
  single-frequency maxima.  That of \texttt{gauss\_D}, Eq.~\eqref{eq:pD}, is the
  deterministic inner step of the two-step composite estimator of
  Table~\ref{alg:twostep}.  The joint template-fit MLE still carries an
  $O(1/N_{\rm total})$ bias with $N_{\rm total}\equiv N_sN_f$, which is why the
  exact-Whittle arm shows $R\le1.002$ rather than exactly $1$.
\end{table*}

\begin{align}
  -2\ln\mathcal{L}^{\rm W}(P|D) &= 2N_s\!\left(\ln P + \frac{D}{P}\right),
    \label{eq:lW} \\
  -2\ln\mathcal{L}^{\rm S}(P|D) &= N_s\!\left(\frac{D-P}{D}\right)^{\!2},
    \label{eq:lS} \\
  -2\ln\mathcal{L}^{\rm Q}(P|D) &= N_s\!\left(\frac{D-P}{P}\right)^{\!2},
    \label{eq:lQ} \\
  -2\ln\mathcal{L}^{\rm D}(P|D) &= N_s\!\left(\frac{D-P}{P}\right)^{\!2}
                                    + 2\ln P,
    \label{eq:lD}
\end{align}
where the full likelihoods are the sums of these single-bin forms over frequency bins and
segments; the \texttt{gauss\_D} likelihood is the contrast arm of this paper.
Its per-bin maximum-likelihood estimator is obtained by maximising the
single-bin objective of Eq.~\eqref{eq:lD} with respect to the model PSD
$P$ at fixed periodogram, which returns the quadratic
$P^2+N_sDP-N_sD^2=0$.  The positive root is
\begin{align}
  \hat P^{\rm D}_k
    &= D_k\,\frac{N_s}{2}\Bigl(\sqrt{1+\frac{4}{N_s}}-1\Bigr)
       \notag\\
    &= D_k\!\left(1-\frac{1}{N_s}+O(N_s^{-2})\right),
  \label{eq:pD}
\end{align}
which is non-negative by construction and differs from the sample mean $D_k$ only through the
log-determinant normalisation term of Eq.~\eqref{eq:lD}: writing the prefactor of
Eq.~\eqref{eq:pD} as $f_2$ and using $\langle D_k\rangle=P_k$, its fractional bias is
$\mathcal{B}=f_2-1=-1/N_s+O(N_s^{-2})$.  The closed form makes $\hat P^{\rm D}_k$ the fast,
deterministic inner step of the two-step estimator of Table~\ref{alg:twostep}, avoiding the
pivot-direction local minimum of a full \texttt{gauss\_D} maximiser.  This per-bin $\mathcal{B}=-1/N_s$ is the small, removable point bias, distinct from the
much larger inflation of the composite two-step estimator, which is an
estimator-approximation effect of the unweighted log-space template fit.

Because $\hat P^{\rm D}_k$ is non-negative it also carries a per-bin rectification absent from
the Whittle likelihood.  The two-step estimator of Table~\ref{alg:twostep} realises
$-2\ln\mathcal{L}^{\rm D}$ computationally.  Its inner step evaluates $\hat P^{\rm D}_k$ per
bin and its outer fit matches $\ln P(\theta;f)$ to $\ln\hat P^{\rm D}_k$ by least squares,
minimising the residual term of Eq.~\eqref{eq:lD}; the $2\sum_k\ln P_k(\theta)$
log-determinant term is not carried, so the procedure is a composite plug-in estimator
whose residual template-fit bias is the subject of Sections~\ref{sec:biasdiag}
and~\ref{sec:vectorbias}.  The
remaining primitives do \emph{not} share it, since maximising each of
Eqs.~\eqref{eq:lW}--\eqref{eq:lQ} with respect to $P$ gives the unbiased sample mean
$\hat P_k=D_k$, as recorded in Table~\ref{tab:likes}.

All periodogram distributions in this work use the one-sided convention
$\langle\hat C_k\rangle = S_h$ with variance $S_h^2/N_s$, the $\chi^2_{2N_s}/(2N_s)$ law, whose
density prefactor $1/(2N_s)$ and variance $1/N_s$ are the same normalisation; the Whittle
likelihood of Eq.~\eqref{eq:whittle}, the averaged-periodogram distribution and the
\texttt{gauss\_D} MLE of Eq.~\eqref{eq:pD} are thereby mutually consistent.

\subsection{The single-parameter diagnostic}
\label{app:prepro}

Table~\ref{tab:prepro} collects the single-parameter diagnostic constructed from
Eqs.~\eqref{eq:bias_fprr} and~\eqref{eq:stat_fprr} over the segmentations of the ground-based
and space-borne networks.  The two bin counts are kept apart as in Sec.~\ref{sec:biasdiag}:
the ratio $\mathcal{B}/\mathcal{S}$ is evaluated on the full-band Nyquist count $N_f^{\rm Nyq}$ and
$\mathcal{S}_{\rm FIM}$ reproduces the Fisher-matrix amplitude uncertainty of
Ref.~\cite{Franciolini:2025PRD}.  At the $11.5\,$d LISA
baseline the Gaussian diagnostic reaches $\mathcal{B}/\mathcal{S}=62.5$, the value that fixes
the reporting scales $R_i$ and $\rho_i$ introduced in Sec.~\ref{sec:biasdiag}.
\begin{table*}[tp]
  \centering
  \caption{The single-parameter diagnostic from $\mathcal{B}=-1/N_s$
  [Eq.~\eqref{eq:bias_fprr}] and $\mathcal{S}=1/\sqrt{N_fN_s}$ [Eq.~\eqref{eq:stat_fprr}],
  with the same likelihood primitives as Table~\ref{tab:likes}.  The $\mathcal{S}$ and
  $\mathcal{B}/\mathcal{S}$ columns use the full-band Nyquist bin count,
  $\mathcal{S}_{\rm FIM}$ the Fisher-matrix amplitude uncertainty of
  Ref.~\cite{Franciolini:2025PRD}.  LVK denotes the LIGO--Virgo--KAGRA
  network and ET the Einstein Telescope.}
  \label{tab:prepro}
  \small\setlength{\tabcolsep}{4.5pt}
  \begin{tabular}{lcccccccc}
    \toprule
    Detector & $T_{\rm seg}$ & $T_{\rm obs}$ & $N_s$
            & $N_f$ & $\mathcal{S}$ & $\mathcal{B}$ & $\mathcal{B}/\mathcal{S}$ & $\mathcal{S}_{\rm FIM}$ \\
    \midrule
    LVK           & $4\,{\rm s}$     & 1.48\,yr & $1.17\!\times\!10^{7}$ & $4.0\!\times\!10^{3}$
            & $4.6\!\times\!10^{-6}$ & $-8.6\!\times\!10^{-8}$ & $0.019$ & $1.4\!\times\!10^{-2}$ \\
    LVK           & $10\,{\rm min}$  & 1.48\,yr & $7.8\!\times\!10^{4}$  & $6.0\!\times\!10^{5}$
            & $4.6\!\times\!10^{-6}$ & $-1.3\!\times\!10^{-5}$ & $2.78$ & --- \\
    ET            & $4\,{\rm s}$     & 0.99\,yr & $7.8\!\times\!10^{6}$  & $4.0\!\times\!10^{3}$
            & $5.7\!\times\!10^{-6}$ & $-1.3\!\times\!10^{-7}$ & $0.023$ & $4.6\!\times\!10^{-5}$ \\
    ET            & $10\,{\rm min}$  & 0.99\,yr & $5.2\!\times\!10^{4}$  & $6.0\!\times\!10^{5}$
            & $5.7\!\times\!10^{-6}$ & $-1.9\!\times\!10^{-5}$ & $3.40$ & --- \\
    LISA          & $11.5\,{\rm d}$  & 4.0\,yr  & $127$                  & $4.968\!\times\!10^{5}$
            & $1.26\!\times\!10^{-4}$ & $-7.87\!\times\!10^{-3}$ & $62.5$ & $4.2\!\times\!10^{-4}$ \\
    Taiji         & $11.5\,{\rm d}$  & 6.0\,yr  & $190$                  & $4.968\!\times\!10^{5}$
            & $1.03\!\times\!10^{-4}$ & $-5.26\!\times\!10^{-3}$ & $51.1$ & $2.8\!\times\!10^{-4}$ \\
    \bottomrule
  \end{tabular}
  \normalsize
\end{table*}

\subsection{The two-step realisation of the contrast arm}
\label{app:twostep}

The likelihood primitives of Table~\ref{tab:likes} are per-bin objectives.  A pipeline has to
choose how to realise them.  The two-step realisation used as the contrast arm of this work is
collected in Table~\ref{alg:twostep}.  Its outer fit is an unweighted log-space fit whose
residuals are homoscedastic, because $\mathrm{Var}[\ln\hat P_k]=1/N_s$ for every member of the
family of Eq.~\eqref{eq:gapblind}.  An inverse-variance reweighting therefore differs from the
unweighted fit by an overall constant and leaves it unchanged.  The per-trial estimate is the
minimiser of
\begin{equation}
  \hat\theta^{(m)} \equiv \operatorname*{arg}\,\operatorname*{min}_{\theta}\;
  \sum_{k} w_k^{(m)}\Bigl[\ln P(\theta;f_k)-\ln\hat P^{(m)}_k\Bigr]^2\,,
  \label{eq:thetahat}
\end{equation}
with the first-stage per-bin estimator entering as the data.

\begin{table*}[tp]
  \centering
  \caption{The two-step realisation of the \texttt{gauss\_D} contrast arm, which extracts the
  point bias and the estimator covariance for a template $P(\theta; f)$ and a given $N_s$.}
  \label{alg:twostep}
  \begin{tabular}{@{}p{0.95\linewidth}@{}}
    \toprule
    \textbf{Input:} true parameters $\theta_0$, number of averaged segments
    $N_s$, number of Monte-Carlo trials $N_{\rm trials}$, frequency grid
    $f_k$. \\
    \textbf{Output:} the point bias $\mathbf{b}$ and the estimator covariance
    $\mathbf{C}_{\hat\theta}$ of Eq.~\eqref{eq:mse}. \\
    \midrule
    \textbf{1. Synthesise trials.} For $m = 1, \dots, N_{\rm trials}$, draw \\
    \qquad$\hat C_k^{(m)} = P(\theta_0; f_k)\, \chi^2_{2N_s}/(2N_s)$. \\
    \textbf{2. Per-bin estimate.} For each trial compute the closed-form
    per-bin estimator of Eq.~\eqref{eq:pD}, \\
    \qquad$\hat P^{(m)}_k = \hat C_k^{(m)} \, \frac{N_s}{2}
    (\sqrt{1+4/N_s}-1)$. \\
    \textbf{3. Outer fit.} Fit $\log P(\theta; f_k)$ to
    $\log \hat P^{(m)}_k$ by L-BFGS-B to obtain
    $\hat\theta^{(m)}$ [Eq.~\eqref{eq:thetahat}]. \\
    \textbf{4. Ensemble statistics.} Compute \\
    \qquad$\bar\theta = \frac{1}{N_{\rm trials}} \sum_m \hat\theta^{(m)}$,
    \quad $\mathbf{b} = \bar\theta - \theta_0$, \\
    \qquad$\mathbf{C}_{\hat\theta} = \frac{1}{N_{\rm trials}-1}
    \sum_m \bigl(\hat\theta^{(m)} - \bar\theta\bigr)
    \bigl(\hat\theta^{(m)} - \bar\theta\bigr)^{\rm T}$. \\
    \bottomrule
  \end{tabular}
\end{table*}

The synthetic data-generating process of the contrast arm
is the Whittle model itself, each fine bin following $P(\theta_0;f)\chi^2_{2N_s}/(2N_s)$,
which is the law the exact-Whittle likelihood assumes.  The contrast arm is therefore an
implementation check that must recover the analytic null of Sec.~\ref{sec:gapfacts}, with the
residual being the $O(1/N_{\rm total})$ sampling shift of the joint fit.  The outer fit is
not a harmless rewriting of the objective.  Evaluating three outer objectives on the same
first-stage estimates and ensembles isolates the two approximations the log-space form
contains: the log-space fit returns a $\log_{10}A$ bias of $+0.375$, the first term of
Eq.~\eqref{eq:lD} alone returns $+0.016$, and restoring in addition the $2\sum_k\ln P_k$
log-determinant term returns $+0.138$.  The three differ only at second order in the per-bin
residual, of scale $(P_k-\hat P_k)/P_k\sim N_s^{-1/2}\simeq9\%$, but the template is strongly
degenerate along the amplitude--shape direction and that degeneracy promotes the second-order
difference to a first-order difference in the recovered bias.  The contrast arm is what an
unweighted log-space outer fit adds to the level statistics.  It is not a property of the
likelihood and, as Sec.~\ref{sec:gapblind} shows, not a property of the estimator efficiency
either.

\section{Network Fisher details: ORF forms, Galactic foreground, and T-channel leakage}
\label{app:network}

\subsection{Adopted ORF forms}
\label{sec:orf}

We adopt closed-form intra-constellation ORFs of \cite{Cornish:2003,Prince:2002,Seto:2006} at
long wavelengths ($f \ll f_*$), matched at high frequencies to the equal-arm envelope
(unnormalised sinc, $\mathrm{sinc}\,x \equiv \sin x/x$, first zero at $x=\pi$):
\begin{align}
  \Gamma_{AA}^{\rm intra}(f)   &= \tfrac{1}{2}\,{\rm sinc}^2(2\pi f L/c), \label{eq:orfdef}\\
  \Gamma_{AE}^{\rm intra}(f)   &= 0, \notag\\
  \Gamma^{\rm inter}_{12}(f)  &= \tfrac{3}{8}\,{\rm sinc}^2(2\pi f d/c)
                                  (1+\tfrac{1}{2}\cos\Delta\phi)
                                  \notag\\
                                &\qquad - \tfrac{1}{4}\,{\rm sinc}^2(\pi f d/c)\cos\Delta\phi,
\end{align}
where ``intra'' refers to channel pairs within one constellation and ``inter'' to pairs across
two.  In the intra term the argument is $f/f_*$ with $f_*=c/(2\pi L)$, $L$ the arm length, and
the envelope vanishes at the equal-arm Michelson null $f=c/(2L)$ ($60\,$mHz for LISA, $50\,$mHz
for Taiji).  In the inter term $d$ is the heliocentric separation of the two constellations and
$\Delta\phi$ their orbital-longitude difference; the envelope oscillates on the separation
scale $c/d\simeq3\,$mHz, with long-wavelength limit $\Gamma^{\rm inter}_{12}(0)=0.344$.  For
LISA and Taiji, $\Delta\phi = 60^\circ$ gives $d=1.50\times10^{11}\,$m, consistent with the
$10^{11}\,$m separation assumed in the LISA--Taiji network literature.  The exact
separated-detector ORF is instead a spherical-Bessel expansion in $y=2\pi fd/c$,
$\gamma^{T}_{ab}=\Theta_1(y,\beta)\mp\Theta_2(y,\beta)$ with $\beta$ the angle between the two
information planes~\cite{SetoTaruya:2007,Wang:2021,Wang:2023,Omiya:2020}, which oscillates
rather than decaying.  We retain the closed-form envelope for analytical transparency, and
Sec.~\ref{sec:network} shows that the 4-ch network ratio moves by only $5\%$ across the
envelope, this Bessel form, the standard plane angles, and setting the inter-constellation term
to zero.

These are the channel overlap-reduction functions in the A/E/T basis, normalised so that
$\Gamma_{AA}(0)=\Gamma_{EE}(0)=1/2$ and $\Gamma_{TT}(0)=0$ for an isotropic background.

\subsection{Galactic confusion foreground}\label{sec:galc}
The largest astrophysical foreground below $\sim10\,{\rm mHz}$ in LISA and Taiji is the
unresolved Galactic population of white-dwarf (WD) binaries, supplemented by the ultracompact
AM\,CVn-type helium-transfer systems.  It is ``dominant'' in the sense of outranking the other
astrophysical foregrounds, not the instrumental noise: as quantified below, its strain power
stays below that of the instrumental noise over $0.3$--$5\,$mHz, peaking near $2$--$3\,$mHz at
about a quarter of it for LISA and about a half for Taiji.  We adopt the
analytical parametrisation of~\cite{Cornish:2017} (lineage: the confusion-noise estimates
of~\cite{Bender:1997} and the LISA-confusion model of~\cite{Crowder:2005}):
\begin{align}
  \Omega_{\rm WD}(f)
    &= A_{\rm WD}\,(f/f_{\rm WD})^{\alpha_{\rm WD}}\,
       \exp\!\bigl[-(f/f_{\rm cut})^{N_{\rm cut}}\bigr]
       \notag\\
    &\qquad \times \bigl[1+\delta_{\rm AM\,CVn}(f)\bigr],
  \label{eq:galc}
\end{align}
with $A_{\rm WD} = 1.5\times10^{-12}$ at $f_{\rm WD} = 1$ mHz, $\alpha_{\rm WD} = 2/3$ (the
chirp-mass spectral index), $f_{\rm cut} = 4$ mHz and $N_{\rm cut} = 2$, the strain PSD
following from Eq.~\eqref{eq:omega2sh} as
$S_h^{\rm gal}(f)=(3H_0^2/4\pi^2)\,\Omega_{\rm WD}(f)f^{-3}$.  The bump is the log-normal
$\delta_{\rm AM\,CVn}=A_{\rm b}\exp[-\ln^2(f/f_{\rm b})/2\sigma_{\rm b}^2]$, with
$A_{\rm b}=0.3$, $f_{\rm b}=3\,{\rm mHz}$ and $\sigma_{\rm b}=1/3$.  This foreground is
comparable to, rather than dominant over, the instrumental noise below a few mHz, raising the
LISA and Taiji noise floors by about $16\%$ and $52\%$ at $3\,$mHz
(Fig.~\ref{fig:bench}a); we therefore treat it as effective noise, which is standard and
slightly conservative.  The global-fit treatment of \cite{Littenberg:2020} recovers $\sim 20\%$
of the SGWB sensitivity in the $1$--$3\,$mHz band.

\subsection{T-channel leakage}
\label{sec:tchannel}

The information is carried by the $A$ and $E$ combinations of each constellation, and the
Sagnac combination $T$ of the three Michelson variables is treated as a GW-null stream.  The
triplet $A,E,T$ is the optimally-combined set identified by~\cite{Prince:2002} (response
function~\cite{Cornish:2003}; review of the TDI construction~\cite{Tinto:2021}).  The three
arm-pair channels $X$, $Y$, $Z$ of an equal-arm equilateral constellation are related by the
$120^\circ$ rotation of the triangle, so for a statistically isotropic, unpolarised background
their signal covariance is circulant and is diagonalised by $A=(Z-X)/\sqrt2$,
$E=(X-2Y+Z)/\sqrt6$, $T=(X+Y+Z)/\sqrt3$
\cite{Prince:2002,Cornish:2003,Seto:2006,Smith:2019}: writing the circulant entries as $G_0$
(diagonal) and $G_1$ (off-diagonal), $\Gamma_{AA}=\Gamma_{EE}=G_0-G_1$,
$\Gamma_{TT}=G_0+2G_1$, while $\Gamma_{AE}=\Gamma_{AT}=\Gamma_{ET}=0$.  The cross-overlaps of
$T$ with $A$ and $E$ therefore vanish identically, because the $120^\circ$ symmetry gives
$\langle XX\rangle=\langle YY\rangle=\langle ZZ\rangle$ and
$\langle XY\rangle=\langle XZ\rangle=\langle YZ\rangle$, so that
$\langle (X-Y)(X+Y+Z)\rangle=0$ and $\langle (X+Y-2Z)(X+Y+Z)\rangle=0$ for any spectral shape.
The $T$--$T$ overlap vanishes at leading order as well, because the three arms telescope at
zero frequency: writing $\mathcal{A}(\hat n)$ for the low-frequency response of the pair of
links along $\hat n$,
$X+Y+Z=[\mathcal{A}_{12}-\mathcal{A}_{13}]+[\mathcal{A}_{23}-\mathcal{A}_{12}]+[\mathcal{A}_{13}-\mathcal{A}_{23}]=0$
for every sky direction and polarisation.  Both properties are exact, so $T$ adds no
information about $\Omega_{\rm GW}$ in this limit.

The cancellation is complete only for equal arms in the limit $fL/c\to 0$.  Averaging the full
TDI response over sky direction and polarisation gives
\begin{equation}
  \frac{\Gamma_{TT}}{\Gamma_{AA}} = \frac{1}{1800}\,\Bigl(\frac{f}{f_*}\Bigr)^{6} \bigl[1+\mathcal O\bigl((f/f_*)^2\bigr)\bigr] ,
  \label{eq:gammatt}
\end{equation}
where the coefficient is the long-wavelength limit of the ratio of the analytic sky- and
polarisation-averaged TDI-$A$ and TDI-$T$ overlap reduction functions
of~\cite{Flauger:2021} (their argument being $f/f_*=2\pi fL/c$, with $f_*=c/2\pi L$); the sixth
power is the residual suppression of a null-stream combination whose arms telescope only at
zero frequency.  The form gives $\Gamma_{TT}/\Gamma_{AA}=1.2\times10^{-11}$ at $1\,$mHz and
$1.9\times10^{-7}$ at $5\,$mHz, the upper edge of the window ($f\lesssim0.3\,f_*\approx6\,$mHz)
where the expansion is valid; beyond it the full TDI response of Eq.~\eqref{eq:orfdef}
applies, and a $1\%$ arm-length mismatch changes it by less than $1\%$.  A
stochastic-background estimator squares the overlap reduction function, so the neglected term
enters at order $(f/f_*)^{12}$.  Switching Eq.~\eqref{eq:gammatt} on changes
$\sigma_{\log_{10}A}$ by less than $3\times10^{-6}$ in relative terms for LISA and LISA+Taiji,
and a deliberately exaggerated flat leakage of $10^{-2}$ changes it by less than
$2\times10^{-5}$; the null-stream treatment therefore does not limit the network forecast.
Setting the response $\mathcal R_T$ to zero in Eq.~\eqref{eq:networkcov} discards this residual.

\bibliographystyle{apsrev4-2}
\bibliography{reference}

\end{document}